\documentclass[aps,pre,twocolumn,superscriptaddress]{revtex4-1}

\usepackage[english]{babel}
\usepackage[utf8x]{inputenc}
\usepackage{amsmath,amsfonts,amssymb,amsbsy,eucal,dsfont}
\usepackage{natbib}
\usepackage{color}
\usepackage[colorlinks]{hyperref}
\usepackage{graphicx}
\usepackage{rotating}

\newcommand{\hide}[1]{}

\begin{document}

\title{Short-time Rheology and Diffusion in Suspensions of Yukawa-type Colloidal Particles}

\author{Marco Heinen}
\email[]{m.heinen@fz-juelich.de}
\affiliation{Institute of Complex Systems (ICS-3), Forschungszentrum J\"ulich, D-52425 J\"ulich, Germany}

\author{Adolfo J. Banchio}
\affiliation{FaMAF, Universidad Nacional de C{\a 'o}rdoba, IFEG-CONICET, Ciudad Universitaria, 5000 C{\a 'o}rdoba, Argentina}

\author{Gerhard N\"agele}
\affiliation{Institute of Complex Systems (ICS-3), Forschungszentrum J\"ulich, D-52425 J\"ulich, Germany}

\date{\today}

\renewcommand{\figurename}{Fig.}
\renewcommand{\figuresname}{Figs.}
\renewcommand{\refname}{Ref.}
\newcommand{\refsname}{Refs.}
\newcommand{\expressionname}{Eq.}
\newcommand{\expressionsname}{Eqs.}
\newcommand{\sectionname}{Sec.}

\begin{abstract}
A comprehensive study is presented on the short-time dynamics in suspensions of charged colloidal spheres. 
The explored parameter space covers the major part of the fluid-state regime, with colloid concentrations extending up to the freezing transition.
The particles are assumed to interact directly by a hard-core plus screened Coulomb potential, and indirectly by solvent-mediated
hydrodynamic interactions (HIs). By comparison with accurate accelerated Stokesian Dynamics (ASD) simulations of the hydrodynamic function $H(q)$,
and the high-frequency viscosity $\eta_\infty$, we investigate the accuracy of two fast and easy-to-implement analytical schemes.
The first scheme, referred to as the pairwise additive (PA) scheme, uses exact two-body hydrodynamic mobility tensors. It is in good agreement with the
ASD simulations of $H(q)$ and $\eta_\infty$, for smaller volume fractions up to about $10\%$ and $20\%$, respectively. The second scheme
is a hybrid method combining the virtues of the $\delta\gamma$ scheme by Beenakker and Mazur with those of the PA scheme.
It leads to predictions in good agreement with the simulation data, for all considered concentrations, combining thus precision with
computational efficiency. The hybrid method is used to test the accuracy of a generalized Stokes-Einstein (GSE) relation proposed by
Kholodenko and Douglas, showing its severe violation in low salinity systems. For hard spheres, however, this GSE relation applies decently well.
\end{abstract}


\maketitle

\section{Introduction}\label{sec:Intro}
Charge-stabilized systems of globular Brownian particles are ubiquitously found over a large range of particle sizes, from large, micron-sized
colloids \cite{Russel1989,RiesePRL2000,HolmqvistPRL2010} down to nanometer-sized proteins \cite{Roosen-Runge2011,Patkowski2005,HeinenBSA2011}.
For many such systems, where van der Waals attractions are to a good approximation negligible, the pair interactions can be described
to good accuracy by a hard-sphere plus repulsive Yukawa (HSY) type pair potential, of range determined by the ionic strength of dissolved
co- and counterions. The HSY model spans the range from neutral hard spheres, corresponding to zero screening length or vanishing Yukawa
potential strength, to long-range electric repulsion occurring in low-salinity systems with large screening length. 
In most applications of the HSY model to charged colloids, the Yukawa-tail of the pair interaction is described by the electrostatic
part of the Derjaguin-Landau-Verwey-Overbeck (DLVO) potential \cite{Verwey_Overbeek1948}.

In experimental data analysis and many theoretical applications, easy-to-implement analytic methods are on demand that allow to
calculate, with good accuracy, short-time dynamic properties, such as diffusion coefficients and high-frequency viscosities with a 
minimal computational effort. Short-time diffusion properties are routinely measured in dynamic light scattering \cite{Pusey1991, heinen2010short},
X-ray photon correlation spectroscopy \cite{RiesePRL2000, BanchioPRL2006}, and neutron spin echo \cite{Farago2003,Longeville2008} experiments. The high-frequency
viscosity $\eta_\infty$ is probed experimentally using torsional viscometers, or on employing approximate generalized Stokes-Einstein (GSE) relations, which 
relate $\eta_\infty$ to a diffusion property \cite{Banchio2008, Mason2010}.

Fast and accurate theoretical methods for calculating dynamic properties are valuable in particular for an extensive data analysis, where different
system parameters such as  concentration,
salt content, particle size and charge, pH-value, and solvent properties, are considered in a broad range of values.

We point out here that short-time dynamic properties are of relevance not only in their own right.
They are also required as input to theories describing colloidal long-time dynamics such as 
mode-coupling and dynamic density functional theory approaches.

A particular challenge in the development of analytic methods is the inclusion of the long-ranged, and for more concentrated systems non-pairwise additive, hydrodynamic
interactions (HIs), which essentially influence the dynamic properties in their short-time behavior. Due to their complexity, the inclusion of HIs
constitutes a severe bottleneck in Brownian dynamics simulations.
The account of many-body HIs in analytic methods is only possible by introducing approximations.
For this reason, it is of prime importance to assess the overall accuracy of analytic methods, by the comparison to precise benchmark results obtained from
computationally elaborate dynamic computer simulations.

In this paper, we discuss the pros and cons of two easy-to-implement analytic methods of calculating short-time dynamic properties, such as the 
wavenumber-dependent hydrodynamic function $H(q)$, and the low shear-rate, high-frequency limiting viscosity $\eta_\infty$.
The first method, referred to as the pairwise additive (PA) scheme, uses exact two-body hydrodynamic mobility tensors including the lubrication part,
but it fully disregards three-body and higher-order hydrodynamic contributions.
The second scheme is a hybrid method combing the virtues of Beenakker and Mazur's so-called $\delta\gamma$-scheme approach for $H(q)$ \cite{Beenakker1983,Mazur1984}
and $\eta_\infty$ \cite{Beenakker1984}, with those of the PA scheme, and precise known results for the special case of neutral hard spheres.
The $\delta\gamma$ scheme accounts for many-body HI contributions in an approximate way. We present the two methods in a self-contained way,
allowing for their easy implementation. Both methods 
require the static structure factor, $S(q)$, or equivalently the radial distribution function (rdf) $g(r)$, as the only input. We calculate
this static input using our recently developed analytic modified penetrating-background corrected rescaled mean spherical approximation (MPB-RMSA)
scheme \cite{Heinen2011,Heinen_Erratum_2011}, which allows for a fast and accurate evaluation of the HSY pair-structure functions.    

The accuracy of both methods for calculating $H(q)$ and $\eta_\infty$ is assessed through comparison with a large number of simulation results, representative
of the full fluid-state regime, which we have obtained using accelerated Stokesian Dynamics simulations. The usefulness of the $\delta\gamma$ scheme
based hybrid method is illustrated by testing the validity of three generalized Stokes-Einstein relations.

The paper is organized as follows. Sec.~\ref{sec:Potential_and_structure} explains the essentials of the HSY model, and the MPB-RMSA
method of calculating $S(q)$. The theoretical background on the short-time dynamics of interacting colloidal particles is included in
Sec.~\ref{sec:General_shorttime}, and Sec.~\ref{sec:Short-time_methods} explains the employed methods of calculating short-time dynamic properties.
Our results for $H(q)$, $\eta_\infty$ and additional related short-time properties are summarized in Sec.~\ref{sec:Results}. Sec.~\ref{sec:GSEs} includes
the test of GSEs, notably that proposed by Kholodenko and Douglas. Our conclusions are given in Sec.~\ref{sec:conclusions}.   

\section{Pair-potential and static structure}\label{sec:Potential_and_structure}

The present study is concerned with charged spherical colloidal particles that interact directly via the
hard-sphere plus Yukawa (HSY) repulsive pair potential
%
\begin{equation}
 \beta u(x) = \left\lbrace
   \begin{array}{ll}
   \infty \,, & x = r/\sigma \leq 1,\\
   \gamma\;\! \dfrac{{e^{-k x}}}{x} \,, & x > 1.
   \end{array}
 \right. \, \label{eq:eff_pair_pot}\\
\end{equation}
The coupling amplitude $\gamma$ and the screening parameter $k$ are given by 
\begin{eqnarray}
\gamma &&= \frac{L_B}{\sigma}\left( \frac{Z e^{k/2}}{1+k/2} \right)^2,\label{eq:DLVOcouplingConst}\\
k^2 &&= \frac{L_B / \sigma}{1 - \phi}\;\! \left( 24 \phi |Z| + 8 \pi n_s \sigma^3  \right).\label{eq:kappa}
\end{eqnarray}
%
This constitutes the repulsive part of the DLVO potential \cite{Verwey_Overbeek1948}.
Here, $\beta = 1/k_B T$, with Boltzmann constant $k_B$, absolute temperature T, colloidal hard-core diameter $\sigma$,
solvent-characteristic Bjerrum length $L_B = \beta e^2/\epsilon$ in Gaussian units, proton elementary charge $e$,
and solvent dielectric constant $\epsilon$. In the DLVO expressions in \expressionsname~\eqref{eq:DLVOcouplingConst} and
\eqref{eq:kappa}, $\gamma$ and $k$ are determined by the number concentration
of monovalent (salt) coions, $n_s$, the effective colloid charge, $Ze$, and the colloid volume fraction, $\phi$.
The square of $k$ is a sum of two contributions, namely
$k^2_c = 24 \phi |Z| L_B / [\sigma(1-\phi)]$ and $k^2_s = 8\pi n_s\sigma^2 L_B / (1-\phi)$. The first one, $k^2_c$,
describes the screening influence of counterions released from the colloid surfaces, and the second one, $k^2_s$, accounts
for the screening influence of monovalent electrolyte ions arising from added salt. As shown in \refsname~\cite{Russel1981,Denton2000},
the factor $1/(1-\phi)$ in \expressionname~\eqref{eq:kappa} corrects for the free volume accessible to the microions
in presence of impermeable colloidal spheres.        

In recent work \cite{Heinen2011,Heinen_Erratum_2011}, we have introduced a semi-analytic Ornstein-Zernike integral equation scheme \cite{Hansen_McDonald1986}
for calculating equilibrium pair-distribution functions, denoted as the
modified penetrating-background corrected mean spherical approximation (MPB-RMSA). This computationally highly effective method
allows for calculating the static structure factor, $S(q)$,
as a function of the scattering wavenumber $q$, for particles interacting by the HSY pair potential given in \expressionname~\eqref{eq:eff_pair_pot}.
The MPB-RMSA is a simple improvement of the PB-RMSA method by Snook and Hayter \cite{Snook1992}, which in turn is based on the frequently
used RMSA method by Hansen and Hayter \cite{Hansen1982}. The analytic simplicity, and the low computational cost of the standard RMSA method is preserved in the MPB-RMSA.
In addition, the MPB-RMSA constitutes a significant improvement over the RMSA by correcting for the RMSA-typical underestimation
of the principal peak height of $S(q)$. 
The excellent accuracy of the MPB-RMSA has been assessed in \refname~\cite{Heinen2011} by means of
extensive parameter studies in comparison to Monte-Carlo (MC) simulations, and results obtained
from the highly accurate, but non-analytic Rogers-Young (RY)
integral equation scheme \cite{Rogers1984}.  
The additional virtue of the MPB-RMSA, which we take advantage of in the present study, is its fast performance.
For a given parameter set $\{\gamma, k, \phi\}$, it allows to compute $S(q)$
in an extended $q$ range in about $0.1$ seconds of cpu time on a desktop PC, which is orders of magnitude faster
than using the RY-scheme, or even more time consuming simulations. 

In the limiting case of neutral hard spheres (HS), attained for $\gamma = 0$ ($Z = 0$) or $k \to \infty$ (very large $n_s$),
the MPB-RMSA reduces to the analytic Percus-Yevick solution \cite{Percus1958, Wertheim1963}, which is known
to give accurate pair-correlation functions for $\phi \lesssim 0.35$.
For larger values $0.35 \lesssim \phi < 0.49$, the Percus-Yevick solution tends to overestimate somewhat the structure
factor peak value, $S(q_m)$, of hard spheres, and to underestimate the rdf contact value $g(x=1^+)$.  

Most of the results on short-time properties discussed in this work are for the parameters $L_B = 5.617$ nm, $\sigma = 200$ nm, and $Z = 100$, 
representative of strongly charged colloidal spheres suspended in an organic solvent.
Numerous MPB-RMSA, RY, and MC simulation results for $S(q)$ and $g(r)$ using these parameters
are included in \figuresname~2, 3 and 4 of \refname~\cite{Heinen2011}. These results cover basically the whole fluid regime, with $\phi$ ranging
from $10^{-4}$ to $0.15$, and $n_s$ from $0$ to $10^{-4}$ M.
For conciseness, and since these results for $S(q)$ and $g(r)$
have been published already, we refrain from replotting them in the present work.

\section{Short-time dynamic properties}\label{sec:General_shorttime}

For characterizing the colloidal short-time regime, one considers the momentum relaxation time, $\tau_B = m/(3\pi\eta_0 \sigma)$,
the time scale $\tau_H=\sigma^2\rho_S/(4 \eta_0)$ of hydrodynamic vorticity diffusion,
and the interaction time $\tau_I = \sigma^2/(4 d_0)$ \cite{Nagele1996, Dhont1996}, where
$m$ is the mass of a colloidal sphere, $d_0 = k_B T/(3\pi\eta_0 \sigma)$ is
the translational free diffusion coefficient for stick hydrodynamic surface boundaries, and 
$\rho_S$ and $\eta_0$ are the mass density and shear viscosity of the suspending Newtonian solvent, respectively.
For a coarse-grained time-resolution where $t \gg \tau_B\sim\tau_H$, the motion of a colloidal particle is erratic and overdamped.
In the present work we focus on the colloidal short-time regime $\tau_B \ll t \ll \tau_I$, during which a particle
has moved a tiny fraction of its size only. This allows for calculating short-time properties using pure equilibrium averages.

Diffusion properties can be measured by a variety of scattering techniques, which commonly determine the dynamic structure factor
\cite{Hansen_McDonald1986},
\begin{equation}
S(q,t) = \lim_\infty \left\langle \frac{1}{N} \sum_{l,j=1}^N \exp\left\lbrace i \mathbf{q}\cdot \left[\mathbf{R}_l(0) - \mathbf{R}_j(t) \right]
         \right\rbrace  \right\rangle , \label{eq:S_of_q_t}
\end{equation}
as a function of scattering wavenumber $q$ and correlation time $t$. The brackets, $<...>$, denote an equilibrium ensemble average.
$N$ is the number of colloid particles in the scattering volume,
$\mathbf{q}$ is the scattering wave vector, and $\mathbf{R}_n(t)$ is the position vector pointing to the center of the $n$-th colloidal particle at time $t$.
Moreover, $\lim_\infty$ denotes the thermodynamic limit $N\to\infty$ and system volume $V\to\infty$, with $n = N/V$ fixed,
which characterizes a macroscopic system. On the colloidal short-time scale, $S(q,t)$ decays exponentially according to \cite{Pusey1991} 
\begin{equation}
\frac{S(q,t)}{S(q)} = \exp \left[ -q^2 D(q) t \right] , \label{eq:S_qt_decay}
\end{equation}
where $D(q)$ is the wavenumber-dependent short-time diffusion function. A statistical-mechanical expression for $D(q)$ follows from the 
generalized Smoluchowski equation in form of the ratio \cite{Jones1991, Nagele1996, Dhont1996}
\begin{equation}
D(q) = d_0 \dfrac{H(q)}{S(q)},\label{eq:H_of_q_definition}
\end{equation}
of the hydrodynamic function
\begin{equation}
H(q) = \lim_\infty \left\langle \frac{k_B T}{N d_0} \sum_{l,j=1}^N \hat{\mathbf{q}} \cdot {\boldsymbol{\mu}^{tt}_{lj}(\mathbf{R}^N)}
       \cdot \hat{\mathbf{q}} \exp \left\lbrace i \mathbf{q}\cdot[\mathbf{R}_l - \mathbf{R}_j] \right\rbrace \right\rangle , \label{eq:H_of_q_microscop}
\end{equation}
and the static structure factor $S(q) = S(q,t=0)$. 
Here, $\hat{\mathbf{q}}$ is the unit vector in the direction of $\mathbf{q}$, and ${\boldsymbol{\mu}^{tt}_{lj}(\mathbf{R}^N)}$ is a translational mobility
tensor linearly relating the hydrodynamic force on a sphere $j$ to the translational velocity of a sphere $l$.
This mobility depends in general on the instantaneous positions, $\mathbf{R}^N$, of
all $N$ particles through the specified hydrodynamic boundary conditions. In this work, stick hydrodynamic boundary conditions
are assumed throughout. The positive-valued hydrodynamic function $H(q)$ is a measure of the influence of HIs on short-time diffusion.
In the (hypothetical) case of hydrodynamically non-interacting particles, $H(q) \equiv 1$, independent of $q$.
The hydrodynamic function can be interpreted
as the reduced short-time generalized mean sedimentation velocity measured in a homogeneous suspension subject to a weak force field collinear with
$\mathbf{q}$ and oscillating spatially as $\cos(\mathbf{q}\cdot\mathbf{r})$. Hence,  
\begin{equation}
\lim_{q\rightarrow0}H(q) = \frac{U_{\text{sed}}}{U_0} \equiv K\label{eq:U_S}
\end{equation}
is equal to the concentration-dependent (short-time) sedimentation velocity, $U_{\text{sed}}$, of a slowly settling suspension of spheres
in units of the sedimentation velocity, $U_0$, at infinite dilution.

The function $H(q)$ can be expressed as the sum, 
\begin{equation}\label{eq:Hq_self_distinct}
H(q)= \dfrac{d_S}{d_0} + H^d(q),
\end{equation}
of a $q$-dependent distinct part, $H^d(q)$, which vanishes for $q\to\infty$, and a self-part equal to the reduced short-time translational self-diffusion
coefficient $d_s/d_0$.
The diffusion coefficient $d_s$ is equal to the short-time slope of the mean-squared displacement, \mbox{$W(t)=1/6 < {[\mathbf{R}(t)-\mathbf{R}(0)]}^2>$},
of a colloidal particle \cite{Dhont1996}.

Two additional diffusion coefficients related
to $D(q)$ are the short-time collective diffusion coefficient $d_c = d_0 K / S(q\to0)$, and the
short-time cage diffusion coefficient $d_{\text{cge}} = D(q_m)$. These two coefficients characterize the decay rates, respectively, of thermally induced concentration
fluctuations of macroscopic wavelengths, and of a wavelength related to the size, $2\pi/q_m$, of the dynamic next-neighbor cage formed around
a particle.   

So far only diffusion properties have been discussed.
A rheological short-time property is the high-frequency limiting viscosity, $\eta_\infty$, which linearly relates the average deviatoric suspension shear
stress to the applied rate of strain in a low-amplitude, oscillatory shear experiment with frequency $\omega \gg 1/\tau_I$.
The statistical-mechanical expression for $\eta_\infty$ is \cite{Cichocki2008}
\begin{equation}\label{eq:etainf_microscop}
\eta_\infty = \eta_0 + \lim_\infty \dfrac{1}{10V} \sum_{\alpha,\beta=1}^3 {\left<\sum_{l,j=1}^N
{{\mu}^{dd}_{lj}}_{\alpha\beta\beta\alpha}(\mathbf{R}^N) \right>} , 
\end{equation}
invoking the Cartesian components, ${{\mu}^{dd}_{lj}}_{\alpha\beta\beta\alpha}$, of the $3 \times 3 \times 3 \times 3$ dipole-dipole mobility
tensor $\boldsymbol{\mu}^{dd}_{lj}$, that relates the symmetric hydrodynamic force dipole moment tensor of sphere $l$ to the rate of strain
tensor related to sphere $j$. For stick hydrodynamic boundary conditions, $\eta_\infty = \eta_0 [ 1 + 2.5\phi + \mathcal{O}(\phi^2)]$, where
$\eta_0$ is the solvent viscosity. The $\mathcal{O}(\phi^2)$ contribution is due to particle interactions. Note here that $<...>$ describes
an equilibrium average with respect to the
unsheared system. Direct interactions affect $\eta_\infty$ only through their influence on the equilibrium particle distribution.

The great difficulty in evaluating \expressionsname~\eqref{eq:H_of_q_microscop} and \eqref{eq:etainf_microscop} to obtain $H(q)$ and $\eta_\infty$,
respectively, lies in the calculation of the hydrodynamic tensors 
${\boldsymbol{\mu}^{tt}_{lj}(\mathbf{R}^N)}$ and $\boldsymbol{\mu}^{dd}_{lj}(\mathbf{R}^N)$, and in the associated many-particle average.
Except for numerically expensive simulations \cite{Banchio2008, BanchioBrady:03, Abade_visco2010, AbadeJCP2010}, it is practically
impossible to gain numerically precise results for concentrated systems. In searching for methods which allow to calculate $H(q)$ and $\eta_\infty$
to decent accuracy with moderate numerical effort, one has to resort to approximate methods.
Two of these methods, namely the so-called pairwise additive (PA) approximation, and the $\delta\gamma$-scheme by Beenakker and Mazur supplemented by
a so-called self-part correction, are discussed in the next sections. 
The methods are presented in a self-contained way to facilitate their implementation by an interested reader. Both methods have in common that
they require $S(q)$, or equivalently $g(r)$, as the only input.
The pros and cons of both methods are assessed in comparison to elaborate computer simulations.

\section{Computational methods}\label{sec:Short-time_methods}

\subsection{Pairwise additive approximation}\label{sec:sub:PA_HI}

In the PA approximation, the $N$-particle translational mobility tensors, ${\boldsymbol{\mu}^{tt}_{lj}(\mathbf{R}^N)}$, 
are approximated by the sum of two-body mobilities according to
\begin{eqnarray}
{\left.{\frac{k_B T}{d_0}\boldsymbol{\mu}^{tt}_{lj}(\mathbf{R}^N)}\right|}_{\text{PA}} =&&~ \delta_{lj} \left[ \mathds{1} +
\sum_{n\neq l}^N \mathbf{a}_{11}(\mathbf{R}_l-\mathbf{R}_n)  \right]\nonumber\\
&& + (1 - \delta_{lj}) \mathbf{a}_{12}(\mathbf{R}_l-\mathbf{R}_j).\label{eq:mob_pa}
\end{eqnarray}
The $3 \times 3$ mobility tensor $\mathds{1} + \mathbf{a}_{11}$ relates,
for an isolated pair of particles in a quiescent fluid, the force on particle 1 to its own velocity.
Correspondingly, $\mathbf{a}_{12}$ relates the force on particle $2$ to the velocity of particle $1$.
The axial symmetry of the two-sphere problem allows to split the reduced mobilities into longitudinal and transverse components,
\begin{equation}
\delta_{ij}\mathds{1} + \mathbf{a}_{ij}(\mathbf{r}) = x_{ij}^a(r)\hat{\mathbf{r}}\hat{\mathbf{r}} + y_{ij}^a(r)\left[ \mathds{1} - \hat{\mathbf{r}}\hat{\mathbf{r}} \right],\label{eq:2bodymob_split}
\end{equation}
where we use the notation from \cite{jeff_oni:84}. The mobility components $x_{ij}^a(r)$ and $y_{ij}^a(r)$
can be expanded analytically in powers of $\sigma/r = 1/x$ using recursion formulas \cite{Schmitz1988}.

In a homogeneous fluid system, the ensemble average of a function $f$ depending on two particle coordinates can be expressed in the thermodynamic limit as
\begin{equation}
\left\langle f(\mathbf{R}_l-\mathbf{R}_j) \right\rangle = \lim_{V\to\infty} \frac{1}{V} \int_V d\mathbf{r}~g(r) f(\mathbf{r}).\label{eq:ensembleavg_gr}
\end{equation}
The combination of \expressionsname~\eqref{eq:H_of_q_microscop}, \eqref{eq:mob_pa}, and \eqref{eq:ensembleavg_gr} leads to the following PA results for $d_S$, 
\begin{equation}\label{eq:ds_PA}
{\left.\dfrac{d_S}{d_0}\right|}_{\text{PA}} = 1 + 8 \phi \int_1^\infty dx~x^2 g(x) \left[ x_{11}^a(x) + 2 y_{11}^a(x) -3\right] ,\\
\end{equation}
and for the distinct part of the hydrodynamic function, 
\begin{eqnarray}\label{eq:Hdistinct_PA}
&& {\left.H^d(y)\right|}_{\text{PA}} = - 15\phi\dfrac{j_1(y)}{y} +\\
&&18\phi\int_1^\infty dx~x h(x) \left[j_0(xy) - \dfrac{j_1(xy)}{xy} + \dfrac{j_2(xy)}{6x^2} \right] + \nonumber\\
&&24 \phi\int_1^\infty dx~x^2 g(x) \tilde{y}_{12}^a(x) j_0(xy) +\nonumber\\
&&24 \phi\int_1^\infty dx~x^2 g(x) \left[\tilde{x}_{12}^a(x)-\tilde{y}_{12}^a(x) \right] \times \left[\dfrac{j_1(xy)}{xy}-j_2(xy)\right].\nonumber 
\end{eqnarray}
Here, $y = q\sigma$ is the diameter-scaled wavenumber, $j_n$ is the spherical Bessel function of first
kind and order $n$, and $h$ is the total correlation function defined by $h(x) = g(x)-1$.

We have introduced here the short-range mobility parts 
\begin{eqnarray}\label{eq:short_range_mobilities}
\tilde{x}_{12}^a(x) &&= x_{12}^a(x) - 3/4 x^{-1} + 1/8 x^{-3},\\
\tilde{y}_{12}^a(x) &&= y_{12}^a(x) - 3/8 x^{-1} - 1/16 x^{-3},
\end{eqnarray}
which include all terms in the series expansion except for the far-field terms up to the dipolar (Rotne-Prager) level, which are subtracted off. 

Analogous to the translational diffusivity tensors, the dipole-dipole mobility tensors are approximated in the PA scheme by their self and distinct
two-body contributions ${\boldsymbol{\mu}^{dd}}^{(2)}_{11}(\mathbf{r})$, 
${\boldsymbol{\mu}^{dd}}^{(2)}_{12}(\mathbf{r})$, respectively. In hydrodynamically semi-dilute suspensions, the high-frequency viscosity of
colloidal spheres at low shear-rate is then obtained from
\cite{Batchelor1972, Russel1984, Banchio2008}
%
\begin{eqnarray}
\frac{\eta_\infty}{\eta_0} &&= 1 + \frac{5}{2}\phi(1+\phi) + 60\phi^2\int_1^\infty dx~x^2 g(x) J(x),\label{eq:etainf_PA}\\
J(x) &&= \dfrac{6}{25\pi\sigma^3}\left[{\boldsymbol{\mu}^{dd}}^{(2)}_{11,\alpha\beta\beta\alpha}(\mathbf{x}) +
          {\boldsymbol{\mu}^{dd}}^{(2)}_{12,\alpha\beta\beta\alpha}(\mathbf{x}) \right],~~~  
\end{eqnarray}
%
where the rapidly decaying two-body shear mobility function, $J(x)$, accounts for the two-body HIs.
For stick hydrodynamic boundary conditions, $J(x) = 15/128~x^{-6} + \mathcal{O}(x^{-8})$.
 
For pair-distances $x > 3$, we use explicit analytic expansions up to $\mathcal{O}(x^{-20})$, given in \refname~\cite{Schmitz1988}
for the two-body mobility functions $x_{ij}^a$ and $y_{ij}^a$, and the leading-order far-field expression, $J(x) = 15/128~x^{-6}$, for the shear mobility function.
Since the expansions in $1/x$ converge slowly at small separations, we employ accurate numerical tables for $x < 3$,
which in particular account for lubrication at near-contact distances.
The tables are based on recursion expressions and a lubrication analysis given in \cite{jeff_oni:84}.
 
Using the zeroth-order concentration-expansion for the rdf of hard spheres given by $g^{\text{HS}}(x) = \Theta(x-1) + \mathcal{O}(\phi)$, with $\Theta$ denoting the
unit step function, we have checked that our PA code precisely reproduces the truncated virial expressions 
\begin{eqnarray}
\frac{d_s^{\text{HS}}}{d_0} &&~=~1 - 1.8315\phi + \mathcal{O}(\phi^2),\label{eq:ds_virial}\\
K^{\text{HS}} &&~=~1 - 6.546\phi  + \mathcal{O}(\phi^2),\label{eq:Sedim_virial}\\
\frac{\eta_\infty^{\text{HS}}}{\eta_0} &&~=~1 + \frac{5}{2}\phi + 5.0023\phi^2 + \mathcal{O}(\phi^3),\label{eq:etainf_virial}
\end{eqnarray}
which were obtained in \cite{Cichocki1999, Cichocki2002, Cichocki2003} using a mobility series expansion including terms up to $\mathcal{O}(x^{-1000})$,
and a lubrication correction.

All PA results for $H(q)$ and $\eta_\infty$ discussed in this paper are based on \expressionsname~\eqref{eq:ds_PA}, \eqref{eq:Hdistinct_PA} and
\eqref{eq:etainf_PA}, with $g(x)$ computed in MPB-RMSA.
Dynamical properties predicted by the PA scheme are exact to linear order in $\phi$.
Thus, the PA scheme is especially well-suited for hydrodynamically, but not necessarily structurally,
dilute systems. Charge-stabilized suspensions at low salinity and concentration, where near-contact
configurations are very unlikely, are prime examples of hydrodynamically dilute, but structurally ordered systems, showing
pronounced oscillations in $S(q)$ and $g(r)$. 

Moreover, the PA scheme can be used to check the accuracy of other approximate schemes, such as the
(self-part corrected) $\delta\gamma$ scheme, in the low concentration regime.
At larger volume fractions, however, and for diffusion properties like $d_c$ and $K$, where
non-pairwise additive HIs are particularly influential, the PA approximation is bound to fail.
Note that, while in the present work particles with stick hydrodynamic boundary conditions are considered, the PA scheme can be easily
generalized to porous particles, and particles with slip-stick boundary conditions, simply by using the corresponding
two-body mobility functions given, e.g., in \cite{Schmitz1988}.

\subsection{$\delta\gamma$-method by Beenakker and Mazur}\label{sec:sub:dg}

Different from the PA scheme, which cannot be applied to concentrated systems, the 
renormalized concentration fluctuation (termed $\delta\gamma$) expansion method of Beenakker and Mazur \cite{Beenakker1984, Mazur1984}
is applicable to fluid-ordered systems also at large values of $\phi$, where three-body and higher-order
HI contributions are important. The $\delta\gamma$ method is an effective medium approach based on a partial resummation of many-body HI contributions. 
While applicable to all $\phi$, its results for $H(q)$ and $\eta_\infty$ reveal moderate inaccuracies
at all concentrations, including the very dilute regime where the PA approach becomes exact.
These inaccuracies can be partially traced back to the approximate expressions of ${\boldsymbol{\mu}^{tt}_{lj}(\mathbf{R}^N)}$
and $\boldsymbol{\mu}^{dd}_{lj}(\mathbf{R}^N)$ used in the derivation of the $\delta\gamma$-scheme where, in particular, lubrication
corrections are disregarded.
Higher order terms in the $\delta\gamma$ expansion require as input static correlation functions of increasing order (pair, triplet, and so on)
with swiftly increasing difficulties in their evaluation.

In the present study, we use the easy-to-implement standard version of the $\delta\gamma$ method for which 
(like for the PA scheme) only $S(q)$ is required as input, with the latter calculated here using the MPB-RMSA scheme.
This is the zeroth-order $\delta\gamma$ approximation regarding $H(q)$, and the second-order $\delta\gamma$ approximation regarding $\eta_\infty$.

The zeroth-order $\delta\gamma$ scheme for $H(q)$ has been applied in the past both to neutral and charged colloidal particles, but the
second-order $\delta\gamma$ scheme for $\eta_\infty$ was used so far for neutral hard spheres only. To our knowledge, the present work provides 
the first test of the $\delta\gamma$ scheme for charged, Yukawa-type particles.   

The zeroth-order $\delta\gamma$-scheme expression for $H(q)$ consists of a microstructure-independent part,  
\begin{equation}\label{eq:ds_deltagamma}
{\left.\dfrac{d_s(\phi)}{d_0}\right|}_{\delta\gamma} =
 \frac{2}{\pi} \int_0^\infty dy {\left[\dfrac{\sin(y)}{y}\right]}^2 \cdot {\left[1 + \phi S_{\gamma_0}(y)\right]}^{-1},
\end{equation}
and a structure factor dependent distinct hydrodynamic function part,
\begin{eqnarray}\label{eq:Hdistinct_deltagamma}
{\left.H^d(y)\right|}_{\delta\gamma} &&= \dfrac{3}{2\pi} \int_0^\infty dy' {\left[\dfrac{\sin(y')}{y'}\right]}^2 \cdot {\left[1 + \phi S_{\gamma_0}(y')\right]}^{-1}\nonumber\\
&&\times \int_{-1}^1 d\mu (1-\mu^2)\left[S(|\mathbf{q}-\mathbf{q'}|)-1\right],  
\end{eqnarray}
where $\mu$ is the cosine of the angle between $\mathbf{q}$ and $\mathbf{q'}$ \cite{Genz1991}.
The function $S_{\gamma_0}(y)$, which should not be confused with the static structure factor,
is given in \cite{Mazur1984, Genz1991}
as an infinite sum of wavenumber-dependent contributions with inter-related scalar coefficients $\gamma_0^{(n)}$, $n=0\ldots\infty$. Numerical results for
$\gamma_0^{(n)}$ obtained from a computation truncated at $n = 5$ have been given in \tablename~1 of the original paper by Beenakker and Mazur \cite{Mazur1984}.
Taking advantage of the nowadays available
computing power, we have been able to extend these earlier computations to more terms with truncations at $n =10$ and $15$.
However, our more accurate results for $\gamma_0^{(n)}$ differ from the original results by Beenakker and Mazur
by no more than $3 \%$, and the differences in $H^d(y)|_{\delta\gamma}$, $d_s(\phi)|_{\delta\gamma}$,
and $\eta_\infty|_{\delta\gamma}$ are negligible for all practical purposes.     

The high-frequency limiting viscosity in the second-order $\delta\gamma$-scheme is given by \cite{Beenakker1984}
\begin{eqnarray}
&&{\left.\dfrac{\eta_\infty}{\eta_0}\right|}_{\delta\gamma} = \dfrac{1}{\lambda_0 + \lambda_2},\label{eq:etainf_deltagamma}\\
&&\lambda_0 = {\left[1 + \frac{5}{2}\phi \tilde\gamma_0^{(2)}\right]}^{-1} = 1 - \frac{5}{2}\phi + \frac{215}{168}\phi^2 + \mathcal{O}(\phi^3),\label{eq:lambda0}\\
&&\lambda_2 = \frac{30 \phi}{4 \pi} {\left[\lambda_0 \tilde\gamma_0^{(2)}\right]}^2 \int_0^\infty dy
              \frac{j_1^2(y/2)\left[S(y/\sigma)-1\right]}{1 + \phi S_{\gamma_0}(y/2)},\label{eq:lambda2}
\end{eqnarray}
where $\tilde\gamma_0^{(2)} = \gamma_0^{(2)}/n = 1 + 167/84\phi + \mathcal{O}(\phi^2)$.

Insertion of the low-concentration form, $S^{\text{HS}}(y) = 1 -24\phi j_1(y)/y + \mathcal{O}(\phi^2)$, of the static structure factor
of neutral hard spheres into \expressionsname~\eqref{eq:ds_deltagamma}, \eqref{eq:Hdistinct_deltagamma}, and \eqref{eq:etainf_deltagamma} gives, after a
straightforward calculation, the first-order virial expansion results  
\begin{eqnarray}
{\left.\frac{d_s^{\text{HS}}}{d_0}\right|}_{\delta\gamma} + \mathcal{O}(\phi^2) &&~=~1 - \frac{131}{56}\phi \approx 1 - 2.339\phi,\\
{\left.K^{\text{HS}}\right|}_{\delta\gamma} + \mathcal{O}(\phi^2)  &&~=~1 - \frac{411}{56}\phi \approx 1 - 7.339\phi,\label{eq:Sedim_virial_dg}\\
{\left.\frac{\eta_\infty^{\text{HS}}}{\eta_0}\right|}_{\delta\gamma}  + \mathcal{O}(\phi^3) &&~=~1 + \frac{5}{2}\phi + \frac{1255}{168}\phi^2\nonumber\\
&&~\approx~1 + \frac{5}{2}\phi + 7.47\phi^2,\label{eq:etainf_virial_dg}
\end{eqnarray}
predicted by the $\delta\gamma$ scheme. The magnitudes of the linear virial coefficients of $d_s^{\text{HS}}/d_0$ and $K^{\text{HS}}$, and of
the quadratic coefficient of $\eta_\infty^{\text{HS}}/\eta_0$, overestimate the precise values given in
\expressionsname~\eqref{eq:ds_virial}, \eqref{eq:Sedim_virial}, and \eqref{eq:etainf_virial} by $28\%$, $12\%$, and $5\%$ respectively. The effect of
HIs on these quantities on the pair-level is thus overestimated by the $\delta\gamma$ scheme. Clearly, the PA scheme is the method of choice when very dilute
systems are considered. We note further that $H^d_{\delta\gamma}(q\to0) = -5\phi + \mathcal{O}(\phi^2)$ for hard spheres, a result quite close to the exact result
of $H^d(q\to0) = - 4.714\phi + \mathcal{O}(\phi^2)$. 
This indicates that the zeroth-order $\delta\gamma$ scheme is in general a better approximation for the distinct part, $H^d(q)$, of the hydrodynamic function
than for its self-part $d_s$.

Interestingly enough, the first-order in $\phi$ result for $H^d(q\to0)$ for hard-spheres predicted by the $\delta\gamma$-scheme, is identical
to the one obtained from the Rotne-Prager (RP)
approximation of the HIs, where only the leading order monopole and dipole terms in the $1/x$ expansion of ${\boldsymbol{\mu}^{tt}_{lj}}$ are retained.
For hard spheres, the first-order virial expansion result, $H(q_m) = 1 - 1.35\phi$, for the principal peak height of $H(q)$
remains valid to high accuracy up to the volume fraction $\phi_f = 0.494$ at freezing \cite{Banchio2008}, whereas in the RP approximation,
peak values of $H(q)$ larger than one are predicted. The main reason for this 
failure of the RP approximation lies in its prediction of $d_s = d_0$ at all $\phi$, whereas the actual $d_s$ of hard spheres
is significantly decreasing with increasing $\phi$, down to the value $d_s^{\text{HS}}(\phi_f) \approx 0.2 d_0$ at freezing.

In low-salinity charge-stabilized systems at low concentrations, where $q_m\sigma \approx 2\pi\sigma n^{1/3} \ll 1$, the $\delta\gamma$-scheme gives predictions
for $H^d(q)$ close to those by the RP approximation. Indeed, these are precisely the systems where the RP approximation can be expected to perform well, explaining
in part the overall success of the $\delta\gamma$ scheme in making reliable predictions for the distinct part of $H(q)$ of charge-stabilized systems.

\subsection{Self-part corrected $\delta\gamma$-scheme}\label{sec:sub:dgCorr}

The key observation regarding the zeroth-order $\delta\gamma$ expression for $H^d(q)$, which depends on $S(q)$ only, is that it
gives overall good results both for neutral and
charged Yukawa-type spheres. In contrast,
the zeroth-order $\delta\gamma$ expansion for $d_s$ in \expressionname~\eqref{eq:ds_deltagamma} depends on $\phi$ only,
independent of the employed pair-potential.  
Comparison with ASD simulation results \cite{Banchio2008}, and experimental data for $d_s$ for charged colloids \cite{Gapinski2006, heinen2010short}, 
show that \expressionname~\eqref{eq:ds_deltagamma} is a decent approximation of $d_s$ for neutral hard spheres only.

The self-diffusion coefficient, $d_s(\phi)$, of charged spheres is in fact larger than the one for neutral spheres at the same $\phi$ \cite{Sinn1999}, since for
the latter near contact configurations are more likely.
Using leading-order far-field mobilities applicable to strongly charged colloids characterized by $q_m \propto \phi^{1/3}$, one finds
for $\phi \lesssim 0.1$ a power-law dependence of $d_s$ according to $d_s/d_0 \simeq 1 - a_t \phi^{4/3}$
\cite{Nagele1996, Nagele1994, Nagele1995, Watzlawek1999, Sinn1999}, differing qualitatively from the regular hard-sphere virial result in
\expressionname~\eqref{eq:ds_virial}. 
The coefficient $a_t \simeq 2.5-2.9$ in the fractional power law varies to a certain extent with the particle size and charge.

For suspensions of strongly charged spheres, where $\phi \lesssim 0.15$, it has been shown \cite{heinen2010short} that 
the PA result for $d_s$ in \expressionname~\eqref{eq:ds_PA} is in better agreement with ASD simulation results, and experimental data, than the corresponding
$\delta\gamma$-scheme result based on \expressionname~\eqref{eq:ds_deltagamma}. For larger concentrations $\phi \gtrsim 0.15$,
the PA scheme overestimates the slowing hydrodynamic influence 
on $d_s$, since it does not account for the shielding of the HIs in pairs of particles by other intervening particles.
Hydrodynamic shielding is a many-body effect which lowers the strength but not the range of the HIs. It should be distinguished from the
screening of HIs by spatially fixed obstacles or boundaries which absorb momentum from the fluid, thereby causing a faster than $1/r$ decay
of the flow perturbation created by a point-like force. The neglect of hydrodynamic shielding (i.e., three-body and higher-order HIs) by the PA scheme
is more consequential for the sedimentation coefficient $K$ than for $d_s$. To the former, the PA scheme is applicable to decent accuracy only
up to $\phi \approx 0.1$ \cite{heinen2010short}, whereas for larger $\phi$, the coefficient $K$ becomes increasingly underestimated. The coefficient $d_s$,
on the other hand, is less
sensitive to the neglect of higher-order HI contributions than $K$ or $d_c$, since the leading-order far-field 
contributions to $x_{11}^a(x)$ and $y_{11}^a(x)$ are of $\mathcal{O}(x^{-4})$, i.e., of shorter range than the leading-order $\mathcal{O}(x^{-1})$
contributions to $K$. 

As a simple improvement over the zeroth-order $\delta\gamma$ scheme for the $H(q)$ of charged particles, which preserves its analytic simplicity, we therefore
use the self-part corrected expression  
\begin{equation}\label{eq:Hq_dgCorr}
{\left.H(y)\right|}_{\delta\gamma_\text{corr}}= {\left.\dfrac{d_s}{d_0}\right|}_{\text{PA}} + {\left.H^d(y)\right|}_{\delta\gamma},
\end{equation}
with $d_s$ according to \expressionname~\eqref{eq:ds_PA} and $H^d(q)$ according to \expressionname~\eqref{eq:Hdistinct_deltagamma}, bearing in mind that
$[d_s/d_0]_{\text{PA}}$ becomes less reliable for $\phi \gtrsim 0.15$.

For the limiting case of neutral hard spheres at larger $\phi$, it is therefore preferential to use in place of $[d_s/d_0]_{\text{PA}}$
the accurate expression,
\begin{equation}\label{eq:ds_HS}
\left.\frac{d_s}{d_0}\right|^{HS} \approx 1 - 1.8315\phi \left(1 + 0.1195\phi - 0.70\phi^2\right) 
\end{equation}
which, for $\phi \leq 0.5$, agrees well with ASD \cite{Banchio2008} and hydrodynamic force multipole \cite{Abade2011} results, with an accuracy better than $3\%$.
Note that the linear and quadratic order coefficients in \expressionname~~\eqref{eq:ds_HS} have been selected identical to the numerically precise
values $-1.8315$ and $-0.219 = -1.8315 \times 0.1195$, for the respective virial coefficients given in \refname~\cite{Cichocki1999}.
We have determined the cubic coefficient in \expressionname~~\eqref{eq:ds_HS} from a
best fit to recent simulation results in \cite{Abade2011} and \cite{Banchio2008}. The coefficient $0.70$
differs somewhat from the coefficient $0.65$ in \cite{Abade2011}, where simulation results only up to $\phi \leq 0.45$ were considered.  

A self-part corrected version of the $\delta\gamma$ scheme for $H(q)$ was used already in earlier applications, where $d_s$ was considered simply as an
adjustable parameter \cite{Genz1991}, or in more recent work determined using elaborate ASD simulations \cite{Banchio2008}.
For practical purposes, however, it is far more convenient to use the analytic $d_s$ corrections in \expressionsname~\eqref{eq:ds_PA}, \eqref{eq:Hq_dgCorr}
and \eqref{eq:ds_HS}. In the present work, numerous ASD simulation results for $H(q)$, $d_s$, and $\eta_\infty$ have been generated
to provide precise benchmarks for assessing the accuracy of the proposed self-part corrected $\delta\gamma$ scheme.   

While the $\delta\gamma$ scheme for $H(q)$ has been used already for charge-stabilized colloids, to our knowledge the application of the
$\delta\gamma$ scheme for $\eta_\infty$ in \expressionsname~\eqref{eq:etainf_deltagamma}-\eqref{eq:lambda2}
was restricted so far to colloidal hard spheres, where 
the predicted values for $\eta_\infty(\phi)$ are in good agreement, for $\phi \lesssim 0.4$, with experiments and simulation data.
In the following section, we are going to assess the performance of the $\delta\gamma$-scheme expression for the $\eta_\infty$ of
charged particles by comparison with ASD simulation data. Our analysis shows that the second-order
$\delta\gamma$ contribution, $\lambda_2(\phi) < 0$, to $\eta_\infty/\eta_0$ is only weakly
dependent on the shape of the static structure factor. Moreover, the zeroth-order contribution, $\lambda_0(\phi) > 0$, which is only dependent on $\phi$,
dominates for small $\phi$ the contribution $\lambda_2$  in magnitude.
As a consequence, the $\delta\gamma$-predicted values for $\eta_\infty$ change only slightly when going from neutral to charged particles, whereas
simulations, and experiments \cite{Bergenholtz2000}, reveal significantly smaller viscosity values in particular at low salinities.
Thus, the $\delta\gamma$-scheme result in \expressionsname~\eqref{eq:etainf_deltagamma}-\eqref{eq:lambda2} applies to neutral hard spheres only.
However, for the interesting case of low-salinity systems, where the viscosity differences to neutral spheres at equal concentration are largest,
the $\delta\gamma$ scheme can be modified (corrected) in an ad-hoc way, according to   
\begin{equation}\label{eq:etainf_dgCorr}
{\left.\dfrac{\eta_\infty}{\eta_0}\right|}_{\delta\gamma_\text{corr}} = 
    1 + \dfrac{5}{2}\phi(1 + \phi) - \dfrac{1}{\lambda_0} + \dfrac{1}{\lambda_0 + \lambda_2}.
\end{equation}
The motivation for this correction follows from the PA expression in \expressionname~\eqref{eq:etainf_PA}: For a low-salinity system
of strongly repelling particles, one has the scaling $r_m \propto \phi^{-1/3}$ for the peak position, $r_m$, of the rdf, and
$g(r\lesssim r_m) \approx 0$. Since $J(x)$ is of $\mathcal{O}(x^{-6})$, for these systems the integral in \expressionname~\eqref{eq:etainf_PA} is 
of $\mathcal{O}(\phi^{3})$. Hence, to quadratic order in $\phi$, $\eta_\infty$ is determined basically by the
microstructure-independent contribution, $1 + 2.5\phi(1+\phi)$, to \expressionname~\eqref{eq:etainf_PA}.  

In \expressionname~\eqref{eq:etainf_dgCorr}, we correct approximately for this limiting behavior of $\eta_\infty$ by subtracting the structure-independent
``self part'', $1/\lambda_0$, from ${[\eta_\infty/\eta_0]}_{\delta\gamma}$, which renders the remainder of $\mathcal{O}(\phi^{2})$ small,
while adding the term $1 + 2.5\phi(1+\phi)$.  
As we will show in the following, the so-corrected $\delta\gamma$ scheme is in
very good agreement with the ASD viscosity data of low-salinity systems, even up to the freezing transition concentration.
We point out here that \expressionname~\eqref{eq:etainf_dgCorr} is restricted in its applicability to the low-salinity regime of strongly
repelling particles. In contrast, the $d_s$-corrected $\delta\gamma$-scheme for $H(q)$ in \expressionname~\eqref{eq:Hq_dgCorr} applies 
to HSY systems for any set of system parameters $\{\gamma,k,\phi\}$, provided $\phi \lesssim 0.15$, including the crossover regime from neutral to deionized, highly charged
particle systems. For neutral hard spheres, the (uncorrected) $\delta\gamma$ scheme for $\eta_\infty$ performs quite well.

The design of a simple,
corrected $\delta\gamma$ scheme which operates well for arbitrary $\{\gamma,k,\phi\}$, including systems of intermediate salinity, is obstructed by
the limited separability of $\lambda_0$ and $\lambda_2$, and by significant many-body HI contributions for more concentrated systems of 
nearly hard-sphere-like particles at high salinity. For small values $\phi \lesssim 0.1$, the PA method can be used to produce reliable 
predictions of $\eta_\infty$, for arbitrary salinities.

\subsection{Accelerated Stokesian Dynamics simulations}\label{sec:sub:ASD}

The simulation data for the $H(q)$ and $\eta_\infty$ in HSY systems explored in this work, have been generated using an
accelerated Stokesian Dynamics (ASD) simulation code. The details of the simulation
method have been explained in Ref. \cite{BanchioBrady:03}.
It allows to simulate short-time properties of a larger number of spheres, typically up to $N = 1000$,
placed in a periodically replicated simulation box, allowing for improved statistics. Since short-time properties
are obtained from single-time equilibrium averages, we have used equilibrium configurations generated using
a Monte-Carlo simulation method for charged spheres and
a Molecular Dynamics algorithm for neutral hard spheres, with the 
many-sphere HIs accounted for using the ASD scheme. The computed hydrodynamic function,
$H_N(q)$, shows a strong system-size dependence, even when $N$ is not small. We therefore extrapolate $H_N(q)$
to the thermodynamic limit using the finite-size scaling correction \cite{Ladd:90,MoSangani1994}, 
\begin{equation}\label{eq:finitesize_Hq}
    H(q) = H_N(q) + 1.76\;\! S(q) \frac{\eta_0}{\eta_\infty(\phi)}\;\! \left(\phi/N\right)^{1/3} \,,
\end{equation}
which, for $q\to0$ and $q\to\infty$, includes the finite-size corrections for $K$ and $d_s$, respectively.
This finite-size correction formula was initially
proposed by Ladd for hard spheres \cite{SegreBehrend:95,Ladd:90,Ladd:95}, and has been
subsequently applied also to charged spheres \cite{Banchio2008} and solvent-permeable particles \cite{AbadeJCP2010}.
As pointed out by Ladd \cite{Ladd:90}, and explained by Mo and Sangani \cite{MoSangani1994}, $\eta_\infty$
is not critically dependent on $N$ so that finite size scaling extrapolation to $N\to\infty$ is not needed (see also \cite{Abade_visco2010}).
The simulation results discussed in following are obtained from averaging over 2000 configurations,
for systems of typically $N=512$ particles.

\section{Results}\label{sec:Results}

\subsection{Diffusion properties of charged particles}\label{sec:sub:Diffusion_prop}

\begin{figure*}
\begin{center}
\hspace{-1em}
\includegraphics[width=.98\textwidth]{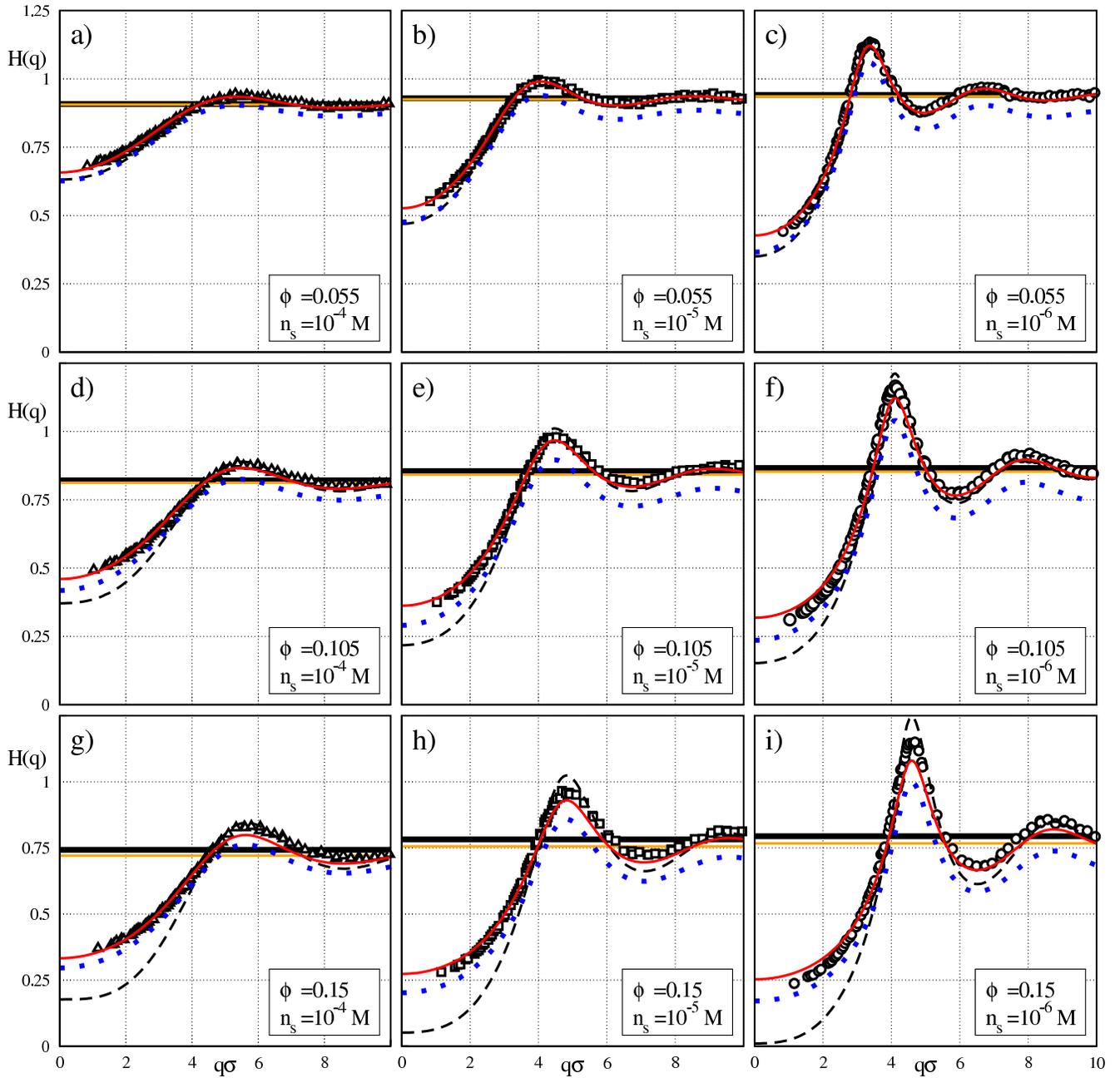}
\end{center}
\vspace{-2em}
\caption{Hydrodynamic function, $H(q)$, for charge-stabilized suspensions of volume fractions, $\phi$, and salt concentrations, $n_s$, as indicated in each panel. 
The panels are ordered with respect to $\phi$, which increases from top to bottom, and $n_s$, which decreases from left to right. 
Open symbols: ASD simulation data. Black dashed, blue dotted, and red solid curves: PA-scheme, $\delta\gamma$-scheme,
and self-part corrected $\delta\gamma$-scheme results, respectively.
Horizontal black and orange lines mark the values for $d_s/d_0$ obtained from ASD simulation and PA-scheme calculations, respectively.
Common system parameters are $L_B = 5.617$ nm, $\sigma =200$ nm, and $Z = 100$.}
\label{fig:Hq_varsalt_various_matrix}
\end{figure*}

In \figurename~\ref{fig:Hq_varsalt_various_matrix}, theoretical and simulation results for $H(q)$ are shown
for nine different systems with common system parameters
$L_B = 5.617$ nm, $\sigma = 200$ nm, and $Z = 100$, representative of suspensions of highly charged colloidal spheres in an organic solvent.
The parameters $\phi$ and $n_s$ are varied, and assume values consisting of all permutations of $\phi = 0.055$, $0.105$, and
$0.15$ and $n_s =10^{-4}$, $10^{-5}$, and $10^{-6}$M. To facilitate the comparison of the different systems, identical axes scales are used
in all nine panels, ordered with respect to $\phi$, which increases from top to bottom, and $n_s$, decreasing from left to right. 
Thus, the strength of the interparticle correlations increases from left to right, and from top to bottom.
\figurename~\ref{fig:Hq_varsalt_various_matrix} serves for analyzing the accuracy of the corrected and uncorrected $\delta\gamma$ schemes, and of the PA scheme,
in comparison to our finite-size corrected ASD results for $H(q)$. 

The results for $H(q)$ obtained by all methods
described in \sectionname~\ref{sec:Short-time_methods}, are included in \figurename~\ref{fig:Hq_varsalt_various_matrix}.
Open symbols represent ASD simulation data, black dashed curves our PA results,
blue dotted curves the zeroth-order $\delta\gamma$ scheme, and red solid lines the self-part corrected $\delta\gamma$ scheme predictions.
The black and orange horizontal lines in each panel mark the reduced short-time self-diffusion coefficient, $d_s/d_0 = H(q\to\infty)$, obtained from
the ASD simulations and the PA scheme, respectively.
As the only input to the three analytic schemes, $S(q)$ and $g(r)$ of each system are obtained using our MPB-RMSA code.
Except for $\phi = 0.105$, the MPB-RMSA results for $S(q)$ and $g(r)$ have been shown already in \figuresname~3 and 4 of \cite{Heinen2011},
and are therefore not included here.
 
The rightmost column of panels in \figurename~\ref{fig:Hq_varsalt_various_matrix} presents results for three systems of strongly charged particles with a very low
residual, but experimentally still accessible, salt content. In the most concentrated system in panel (i), where $\phi = 0.15$ and $n_s = 10^{-6}$ M,
a structure factor peak value $S(q_m) \approx 2.8$ is attained according to the MC simulations. The very same peak value is predicted by the MPB-RMSA and RY
integral equation schemes.
According to the empirical Hansen-Verlet freezing rule, the system in panel (i) is pretty close to the
freezing transition point \cite{StevensRobbins1993,HansenVerlet1969,KremerGrest1986}.

The screening parameter $k$ defined in \expressionname~\eqref{eq:kappa} assumes rather low values of $2.67$, $3.24$, and $3.68$ for the systems in 
panels (c), (f), and (i), with $k^2_c/k^2_s = 1.1$, $2.1$, and $3.0$, respectively.
The relatively large values for $k^2_c/k^2_s$ in these systems indicate that salt microions
contribute little to the electrostatic screening, which instead is dominated by colloid-surface released counterions. 
Different from neutral hard spheres, where $H(q_m)$ decreases linearly with increasing $\phi$, the hydrodynamic function peak heights of the three
low-salinity systems depends non-monotonically on $\phi$, with ASD values $H(q_m) = 1.13$, $1.17$, and $1.15$ 
for the systems in panels (c), (f), and (i), respectively. Such a non-monotonic $\phi$-dependence of $H(q_m)$ is typical for low-salinity systems, as discussed
in \cite{Banchio2008} and \cite{Gapinski2010}. The ASD results for the reduced self-diffusion coefficient, $d_s/d_0$, and the corrected 
$\delta\gamma$-scheme results for the sedimentation coefficient $K$
in panels (c), (f), and (i), follow closely the concentration-scaling predictions $d_s/d_0 = 1 - a_t \phi^{4/3}$ and $K = 1 - a_s \phi^{1/3}$
given in \cite{heinen2010short}, with $a_t = 2.63$ and with $a_s = 1.44$.
The fractional exponents $4/3$ and $1/3$ can be traced back to the $\phi^{1/3}$ scaling in low-salinity systems, of
the structure factor peak position, $q_m$. Note here that $H(q)$ has its principal peak at a wavenumber
practically identical to the peak position of $S(q)$. 

We proceed in our discussion with the systems in the leftmost
column of panels. The salt concentration, $n_s =10^{-4}$ M, of these systems is so large that the neutral hard-sphere (HS) limit is practically reached.
The comparison with the $H(q)$ of genuine neutral hard spheres 
at volume fractions equal to those in panels (a), (d), and (g), shows relative differences of less than $6\%$ in all three cases.
The proximity to genuine neutral hard-sphere systems is manifest also in the large values,
$k = 18.5, 18.6,$ and $18.7$, of the screening parameter, and in the small ratios $k^2_c/k^2_s = 0.01, 0.02,$ and $0.03$, for the systems in
panels (a), (d), and (g), respectively.    

For the smallest considered concentration $\phi=0.055$ (top row of panels in \figurename~\ref{fig:Hq_varsalt_various_matrix}), the differences in the respective $H(q)$
predicted by the analytic methods and the ASD simulations are very small. Since the PA scheme becomes exact at low $\phi$,
this illustrates that, despite their overall inaccuracies,
the self-part corrected, and even the uncorrected $\delta\gamma$-scheme, can be used to obtain good estimates of $H(q)$ also for more dilute suspensions.    

With increasing $\phi$, pronounced differences are observed in \figurename~\ref{fig:Hq_varsalt_various_matrix} between
the PA-scheme and ASD results for $H(q)$. This reflects the expected failure of the PA scheme in concentrated suspensions, where three-body and higher-order
HI contributions become influential. The deviations of the PA-scheme results from the precise simulation data are most pronounced for 
the peak value $H(q_m)$, which is overestimated by the PA scheme,
and in the sedimentation coefficient, $K = H(q\to0)$, which is underestimated. In fact, for the system in panel (i),
the PA prediction for $K$ is just barely larger than zero, turning to unphysical negative values when the volume fraction
surpasses $\phi = 0.154$ at a fixed $n_s = 10^{-6}$ M. However, the PA-scheme values for $d_s/d_0$ remain in very good agreement with the ASD
results, with a relative deviation of less than $3.5\%$ even at $\phi = 0.15$. The values for $d_s/d_0$ predicted by the
uncorrected $\delta\gamma$ scheme are generally in less good agreement with the simulation data, clearly revealed in \figurename~\ref{fig:Hq_varsalt_various_matrix}
by the large-$q$ offset of the corresponding $H(q)$. 

The self-part corrected $\delta\gamma$-scheme results for $H(q)$ in \figurename~\ref{fig:Hq_varsalt_various_matrix} (red solid lines)
illustrate that this hybrid scheme combines the good accuracy of the PA scheme regarding $d_s$, and of the $\delta\gamma$ scheme regarding $H^d(q)$.
Indeed, the corrected $\delta\gamma$-scheme results for $H(q)$ are in overall good agreement with the ASD simulation data for all considered systems,
with the largest deviation of $6\%$ for $H(q_m)$ observed in panel (i).

In closing our discussion of \figurename~\ref{fig:Hq_varsalt_various_matrix}, a short comment is in order regarding the computational cost
caused by the considered methods of computing $H(q)$. The fast and accurate evaluation of $S(q)$ and $g(r)$ by the MPB-RMSA method, in combination with the 
easily evaluable integrals in \expressionsname~\eqref{eq:ds_PA}, \eqref{eq:Hdistinct_PA}, and \eqref{eq:Hdistinct_deltagamma}, has 
allowed us to implement a convenient graphical user interface code, running on a standard desktop PC.
Using this code, MPB-RMSA results for $S(q)$ and $g(r)$, and PA-, $\delta\gamma$-, and self-part corrected $\delta\gamma$-scheme results for $H(q)$, are
obtained in less than $1$ second of cpu time, for a given set of input parameters $\{L_B,\sigma, Z, n_s, \phi\}$.
Thus, all curves depicted in \figurename~\ref{fig:Hq_varsalt_various_matrix},
except for the ASD simulation data, have been obtained altogether in less than a minute on a standard desktop PC.
In comparison, the computation of just one of the computer simulation curves in \figurename~\ref{fig:Hq_varsalt_various_matrix}
required on a standard desktop PC typically 5 hours of cpu time for generating 2000 equilibrated configurations (using our MC method) and approximately 8 hours
of cpu time for computing $H(q)$ with the ASD scheme.
The overall accuracy and fast performance of the hybrid $\delta\gamma$ scheme in \expressionsname~\eqref{eq:Hq_dgCorr}-\eqref{eq:etainf_dgCorr}
make this scheme well-suited for the real-time fitting, even of large sets of experimentally
recorded data for $H(q)$ and $D(q)$ \cite{Holmqvist2011}. 

\subsection{Hybrid $\delta\gamma$ scheme applied to neutral hard spheres}\label{sec:sub:HS_dgcorr}

\begin{figure}
\begin{center}
\hspace{-1em}
\includegraphics[width=.33\textwidth,angle=-90]{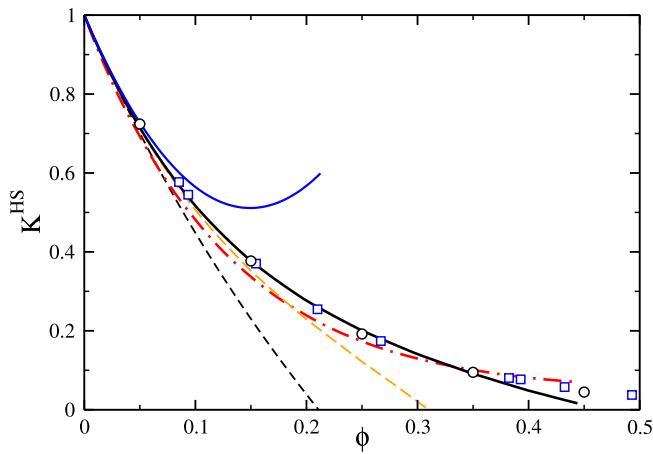}
\end{center}
\vspace{-1em}
\caption{Reduced short-time sedimentation coefficient, $K^{\text{HS}}$, of neutral colloidal hard spheres.
Open circles: Hydrodynamic force multipole simulation data by Abade \textit{et al.} \cite{AbadeJCP2010}.
Open Squares: Lattice-Boltzmann simulation data by Segr\`e \textit{et al.} \cite{SegreBehrend:95}.
Black dashed line: PA-scheme result.
Dashed-dotted red line: uncorrected $\delta\gamma$-scheme result.
Dashed orange line: self-part corrected $\delta\gamma$-scheme result, with $d_s/d_0$ taken from the PA-scheme.
Solid black line: self-part corrected $\delta\gamma$-scheme result, with $d_s/d_0$ according to \expressionname~\eqref{eq:ds_HS}.
Solid blue line: second-order virial result $K^{\text{HS}} = 1 - 6.546\phi + 21.918\phi^2$ \cite{Cichocki2002}.
The static structure factor input was calculated using the analytic Percus-Yevick solution.}
\label{fig:Sedim_HS_Sim_and_var_theo}
\end{figure}

The main virtue of the $d_s$-corrected (zeroth-order) $\delta\gamma$ scheme lies in its good applicability to
charge-stabilized systems. However, it is interesting to assess in more detail its performance in the limiting case of neutral hard
spheres, in particular when the values of $H(q)$ at $q = 0$ and $q = q_m$ are considered. Recall for hard spheres that the accurate expression
for $d_s^{\text{HS}}$ in \expressionname~\eqref{eq:ds_HS}
should be preferentially used instead of the approximate PA result for larger $\phi \gtrsim 0.15$. For neutral spheres, 
higher-order HI contributions to $d_s$ begin to matter at somewhat smaller concentrations than for charge-stabilized particles, where
near-contact configurations are unlikely.

In \figurename~\ref{fig:Sedim_HS_Sim_and_var_theo}, numerically precise Lattice Boltzmann \cite{SegreBehrend:95} and hydrodynamic
force multipole simulation results \cite{AbadeJCP2010} for $K^{\text{HS}}(\phi)$ are compared with the predictions of 
all considered analytical schemes. For $\phi \lesssim 0.35$, the uncorrected $\delta\gamma$ scheme underestimates the simulation data,
showing the opposite trend of a slight overestimation for $\phi \gtrsim 0.35$.
The corrected $\delta\gamma$ scheme with $d_s$-input according to \expressionname~\eqref{eq:ds_HS}, on the other hand,
is in excellent agreement with the simulation data up to $\phi \approx 0.4$,
reflecting the accuracy of the $\delta\gamma$-scheme predictions for $H^d(q)$ also for neutral spheres.

For large volume fractions $\phi \gtrsim 0.4$, however, the distinct part, $K^{\text{HS}} - d_s^{\text{HS}}/d_0$,
of the  sedimentation coefficient is considerably underestimated by the $\delta\gamma$ scheme, to such an extent that the self-part corrected
$\delta\gamma$ scheme prediction for $K^{\text{HS}}$ assumes unphysical
negative values for $\phi \gtrsim 0.45$. Up to $\phi \approx 0.2$, the corrected $\delta\gamma$-scheme prediction for $K^{\text{HS}}$, with $d_s^{\text{HS}}$ obtained
by the PA scheme, lies closer to the simulation data than the uncorrected $\delta\gamma$-scheme result. This can be explained by the precise account
of (two-body) lubrication effects in the PA-scheme, which are not included in the uncorrected $\delta\gamma$ scheme.
At larger concentrations, however, the corrected $\delta\gamma$ scheme, with PA input for $d_s$, increasingly underestimates the sedimentation coefficient
up to the point that, for $\phi \gtrsim 0.31$, unphysically negative values for $K^{\text{HS}}$ are attained. This is a consequence of the already
noted many-sphere hydrodynamic
shielding effect, disregarded in the PA scheme, which lowers the strength of the HIs without reducing their range, leading to a larger self-diffusion
coefficient than predicted on basis of hydrodynamic pair-interactions alone. The neglect of shielding effects by the PA scheme, both in the self- and distinct
parts of $K^{\text{HS}}$, is the reason for the crossover of the PA curve of $K^{\text{HS}}$ to negative values already at $\phi \approx 0.21$.

Regarding again \figurename~\ref{fig:Sedim_HS_Sim_and_var_theo}, we finally note that the second-order virial result,
$K^{\text{HS}} = 1 - 6.546\phi + 21.918\phi^2 + \mathcal{O}(\phi^3)$, derived in \cite{Cichocki2002}
ceases to be applicable for $\phi \gtrsim 0.15$, where its curve bends up to larger values. This is the reason why this second-order virial result
cannot be used, different from the corresponding virial results for $d_s^{\text{HS}}$, $\eta_\infty^{\text{HS}}$, and $H^{\text{HS}}(q_m)$, to construct
analytic extrapolation formulas, valid for all concentrations up to the freezing transition.           

\begin{figure}
\begin{center}
\hspace{-1em}
\includegraphics[width=.33\textwidth,angle=-90]{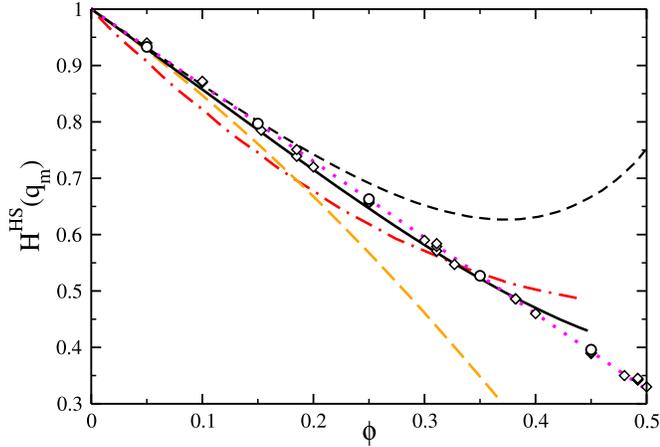}
\end{center}
\vspace{-1em}
\caption{Hydrodynamic function peak value, $H(q_m)$, of neutral hard spheres.
Open circles: Hydrodynamic force multipole simulation results by Abade \textit{et al.} \cite{AbadeJCP2010}.
Open diamonds: ASD simulation data \cite{Banchio2008}.
Black dashed line and dashed-dotted red line: PA-scheme and uncorrected $\delta\gamma$-scheme results, respectively.
Dashed orange line: self-part corrected $\delta\gamma$-scheme results, with $d_s/d_0$ taken from PA-scheme calculations.
Solid black line: self-part corrected $\delta\gamma$-scheme results, with $d_s/d_0$ according to \expressionname~\eqref{eq:ds_HS}.
Dotted curve in magenta: $1 - 1.35\phi$.
The static structure factor input was obtained using the analytic Percus-Yevick solution.
}
\label{fig:Hmax_HS_Sim_and_var_theo}
\end{figure}
We proceed by discussing the concentration dependence of the peak value, $H^{\text{HS}}(q_m)$, of the hydrodynamic function of neutral spheres.
As noted in Sec.~\ref{sec:General_shorttime}, $H(q_m)$ is related to the short-time
cage diffusion coefficient, $d_{\text{cge}} = d_0 H(q_m)/S(q_m)$, characterizing the initial decay rate of density fluctuations of wavelength equal to the
next-neighbor cage size.
\figurename~\ref{fig:Hmax_HS_Sim_and_var_theo} displays the decline of the hard-sphere $H(q_m)$ with increasing $\phi$. To excellent accuracy
up to the freezing volume fraction, this decline is described by the first-order virial result \cite{Banchio2008}    
\begin{equation}\label{eq:Hmax_HS}
H^{\text{HS}}(q_m) = 1 - 1.35\phi. 
\end{equation}
Indeed, all the depicted ASD \cite{Banchio2008} and hydrodynamic force multipole \cite{AbadeJCP2010} values for $H(q_m)$ follow this line, indicating
that, for a so far unknown reason, all higher order virial contributions cancel out. According to \figurename~\ref{fig:Hmax_HS_Sim_and_var_theo},
the uncorrected $\delta\gamma$ scheme significantly underestimates $H^{\text{HS}}(q_m)$ for $\phi \lesssim 0.35$, overestimating it
instead for $\phi \gtrsim 0.4$. In contrast, the corrected $\delta\gamma$-scheme result with the precise $d_s^{\text{HS}}$ according to
\expressionname~\eqref{eq:ds_HS}, is distinctly more accurate in that it only very slightly underestimates the linear decay in
\expressionname~\eqref{eq:Hmax_HS} for
$\phi \lesssim 0.4$. Moreover, for $\phi > 0.4$, the positive-valued deviations from $1 - 1.35\phi$ are substantially smaller than those of the
uncorrected $\delta\gamma$ scheme.

The corrected $\delta\gamma$-scheme prediction for $H^{\text{HS}}(q_m)$, with $d_s$ calculated using the PA scheme,
is a decent approximation up to $\phi \lesssim 0.15$. Its bending over to smaller values occurs for $H^{\text{HS}}(q_m)$ at somewhat larger $\phi$
than in the sedimentation case, indicating that $\delta\gamma$-scheme results for $H^d(q)$ are more accurate at $q = q_m$ than at $q \approx 0$.
The curve for $H^{\text{HS}}(q_m)$ predicted by the PA scheme bends over to larger values at a concentration $\phi \approx 0.37$ vastly
beyond its range $(\phi \lesssim 0.1)$ of applicability. The total neglect in the PA scheme of many-body HIs beyond the pair level implies, at
larger $\phi$, an underestimation
of $d_s$, but to a larger extent an overestimation of $H^d(q_m)$. As a net result, $H^{\text{HS}}(q_m)$ at large $\phi$ is strongly overestimated by the PA scheme.

In summarizing our discussion of hard-sphere systems,
the key message conveyed by \figuresname~\ref{fig:Sedim_HS_Sim_and_var_theo} and \ref{fig:Hmax_HS_Sim_and_var_theo} is
that the corrected $\delta\gamma$ scheme, with $d_s$ according to \expressionname~\eqref{eq:ds_HS},
describes $H^{\text{HS}}(q)$ quite precisely for $\phi \lesssim 0.4$. It can be applied to reasonable accuracy even to larger $\phi$ values,
with the exception of small $q$ values.         

\subsection{High-frequency viscosity}\label{sec:sub:eta_inf}

\begin{figure}
\begin{center}
\hspace{-1em}
\includegraphics[width=.37\textwidth,angle=-90]{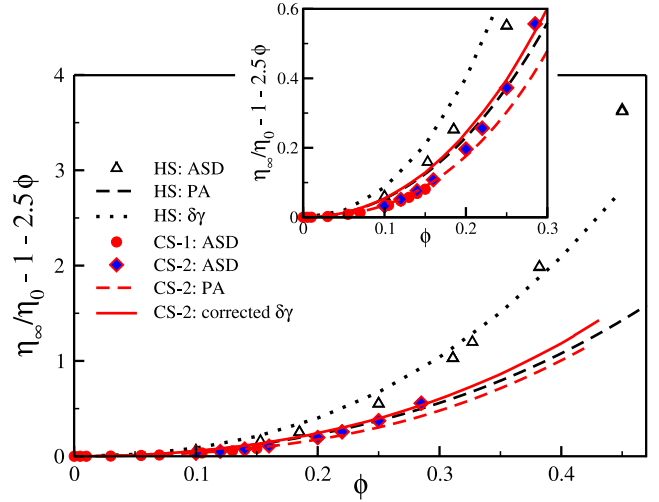}
\end{center}
\vspace{-1em}
\caption{Reduced high-frequency viscosity, $\eta_\infty/\eta_0$, as a function of $\phi$, for a suspension of neutral hard spheres (HS, in black), and 
two deionized charged-sphere suspensions (CS-1 and CS-2, in red).
The leading-order Einstein contribution, $1 + 2.5\phi$, is subtracted off to expose the differences.
Symbols: ASD simulation results. Dashed, dotted and solid lines: PA-, $\delta\gamma$-, and self-part corrected
$\delta\gamma$-scheme results, respectively. All analytic schemes use the MPB-RMSA $S(q)$ as input. The CS-1 viscosity results
represented by red filled circles are ASD data for $L_B =5.617$ nm, $\sigma =200$ nm, and $Z =100$.
The ASD data for the more weakly charged, smaller particles of system CS-2, with $L_B =0.71$ nm, $\sigma =50$ nm, and $Z =70$
are indicated by red diamonds filled in blue. The parameters of system CS-2 have been used in the analytic calculations.
The inset magnifies the details at lower $\phi$.}
\label{fig:eta_inf_HS_CS_ASD_PA_dg_dgCorr}
\end{figure}

In the following, we compare viscosity results obtained by the various methods described in Sec.~\ref{sec:Short-time_methods},
for the two limiting cases of deionized (low-salinity) charged-sphere and neutral hard-sphere suspensions. Results for systems
with intermediate added salt are bracketed by these two limiting cases.

In \figurename~\ref{fig:eta_inf_HS_CS_ASD_PA_dg_dgCorr}, we display our viscosity results for neutral hard spheres (HS) with those for
two deionized suspensions of highly charged spheres (CS) where $n_s = 0$.
Results by all the methods in \sectionname~\ref{sec:Short-time_methods} are shown.
We point out that, in addition to the CS system of \figurename~\ref{fig:Hq_varsalt_various_matrix} with parameters $L_B = 5.617$ nm, $\sigma = 200$ nm, and $Z = 100$
(referred to here as system CS-1), whose ASD results for $\eta_\infty$ in \figurename~\ref{fig:eta_inf_HS_CS_ASD_PA_dg_dgCorr} are indicated by
filled red circles, we additionally show 
viscosity results for another zero-salt system where $L_B = 0.71$ nm, $\sigma = 50$ nm, and $Z = 70$ (labeled CS-2), whose ASD data for $\eta_\infty$ are represented by
red diamonds filled in blue.     
The reason for including in \figurename~\ref{fig:eta_inf_HS_CS_ASD_PA_dg_dgCorr} results for two different
deionized systems, is that system CS-1 freezes at $\phi \approx 0.15$,
whereas systems CS-2 stays fluid up to $\phi \sim 0.3$, allowing us to test the predictions of our analytical methods in a more extended volume fraction range.
The ASD simulation data for $\eta_\infty(\phi)$ for systems CS-1 and CS-2 merge continuously, overlapping nearly perfectly within $0.1 < \phi < 0.15$. This 
indicates that the limiting behavior of $\eta_\infty$ for highly correlated charged spheres is practically reached in both systems.
Therefore, the depicted CS-results for $\eta_\infty$ were calculated in the analytic schemes using the parameters of system CS-2 only.
The deviations in $\eta_\infty$ for the two CS systems are minuscule in all considered analytic schemes.

Consider first the performance of the PA scheme. By its definition, the PA result for $\eta_\infty$ is in very good agreement with the ASD simulation
data at low $\phi$, but the agreement becomes poorer with increasing volume fraction. While the deviations between the ASD and PA results for 
low-salt systems are very small up to $\phi \lesssim 0.2$, for neutral spheres significant differences are visible already for $\phi \gtrsim 0.1$.
These differences originate from the fact that in charged-sphere suspensions, near-contact configurations are disfavored by the electric repulsion,
i.e., charged-sphere systems are hydrodynamically more dilute than neutral sphere suspensions. Since higher-order HI effects on
$\eta_\infty$ in low-salinity charged systems are weaker, for many such systems (including system CS-1), which freeze already at $\phi \lesssim 0.2$,
the accuracy of the PA-scheme for $\eta_\infty$ is sufficiently good in the whole fluid regime.
Regarding $H(q)$, however, the PA-scheme predictions for charged spheres deviate significantly from
the ASD data already at $\phi = 0.105$ (see again \figurename~\ref{fig:Hq_varsalt_various_matrix}).
At larger $\phi$, and in contrast to the ASD data, the PA scheme predicts only a slight enlargement of $\eta_\infty$ in going from charged to neutral spheres.
The distinctly larger values of $\eta_\infty$ for concentrated hard-sphere suspensions are thus mostly due to near-field, many-body HIs which enlarge
the viscous dissipation. Overall, however, $\eta_\infty$ is rather insensitive to the range of the pair potential, at least in comparison to the static
(zero frequency) viscosity which for concentrated systems can become very much larger than $\eta_\infty$ \cite{Banchio1999Gsess}.

The self-part modified $\delta\gamma$ scheme for $\eta_\infty$, defined by \expressionname~\eqref{eq:etainf_dgCorr}, agrees overall very well with the ASD data for
charged spheres in the whole fluid-state concentration regime. Small deviations from the simulation data are noticed at low $\phi$ values only.
Regarding neutral hard spheres, a similar observation applies to the unmodified second-order $\delta\gamma$ scheme, which describes the ASD simulation
data quite well up to $\phi \approx 0.4$.
The slight overestimation of $\eta_\infty^{\text{HS}}$ at lower $\phi$
can be attributed to the non-exact treatment of two-body HI contributions by the $\delta\gamma$ scheme.   

Overall, the high-frequency viscosity of charged-sphere systems at low salinity is well captured by the modified $\delta\gamma$-scheme
in \expressionname~\eqref{eq:etainf_dgCorr}, and for neutral hard spheres 
by the unmodified $\delta\gamma$ scheme (up to $\phi \approx 0.4$). Different from the self-part corrected $\delta\gamma$ scheme for $H(q)$,
which makes reliable predictions for arbitrary salinities, the modified $\delta\gamma$ scheme for $\eta_\infty$ applies to low-salinity
systems only, and the unmodified $\delta\gamma$ scheme only to neutral hard spheres. The reasons for this have been discussed already in subsection \ref{sec:sub:dgCorr}. 
         
Before closing our discussion of $\eta_\infty$, it is of interest to compare the numerical efforts required by the employed methods.
The computation of the 45 ASD data points
for neutral and charged spheres included in \figurename~\ref{fig:eta_inf_HS_CS_ASD_PA_dg_dgCorr} required
about 500 hours of cpu time on a modern desktop PC.
This large time investment should be compared to the few minutes computation time on a comparable PC which were required for the 
results by all considered analytic schemes, amounting to more than one thousand data points on a dense mesh of $\phi$ values.   

\section{Relation between viscosity and diffusion properties}\label{sec:GSEs}

Having quite accurate analytic methods for short-time properties at our disposal, we are in the position to analyze possible relations
between these properties in the whole fluid-state concentration regime. Specifically,
we want to test the validity of two generalized Stokes-Einstein (GSE) relations,
\begin{equation}\label{eq:Self_Cage_GSE}
\frac{D^*(\phi)}{d_0} \times \frac{\eta_\infty(\phi)}{\eta_0} \approx 1 ,
\end{equation}
for $D^*(\phi) = d_s(\phi)$ and $D^*(\phi) = d_{\text{cge}}(\phi)$. In addition, we probe the validity of the Kholodenko-Douglas GSE (KD-GSE) 
relation \cite{Kholodenko1995},  
\begin{equation}\label{eq:KD-GSE}
\frac{d_c(\phi)}{d_0}\times\frac{\eta_\infty(\phi)}{\eta_0}\times\sqrt{S(q\to0, \phi)} \approx 1 ,
\end{equation}
between the collective diffusion coefficient, $d_c = K/S(q\to0)$, $\eta_\infty$, and the square root of the isothermal
osmotic compressibility given by $S(q\to0)$.
In particular the KD-GSE relation has been used in various biophysical and soft matter studies \cite{Gaigalas1995, Cohen1998, Boogerd2001, Nettesheim2008}.

All three considered GSE relations are exact at $\phi = 0$ only. The approximate validity of a GSE relation in concentrated systems is an important issue
in microrheological studies, since this allows to infer a rheological property more easily from a diffusion measurement.
For testing the GSE relations in \expressionsname~\eqref{eq:Self_Cage_GSE} and \eqref{eq:KD-GSE},
we consider here again the two limiting HSY cases of a low-salinity charge-stabilized system and neutral spheres, since the differences 
in the respective short-time dynamic properties are here largest.   

For a precise test of the GSE relations in the case of hard spheres, we take advantage of simple analytic expressions available for all short-time properties
appearing in \expressionsname~\eqref{eq:Self_Cage_GSE} and \expressionname~\eqref{eq:KD-GSE}, with the exception of $K$, for which we use the quite accurate self-part
corrected $\delta\gamma$-scheme result depicted in \figurename~\ref{fig:Sedim_HS_Sim_and_var_theo}. The analytic expressions for hard spheres,
which apply to excellent accuracy up to $\phi = 0.5$, are \expressionname~\eqref{eq:Hmax_HS} for $H^{\text{HS}}(q_m)$, \expressionname~\eqref{eq:ds_HS} for
$d_s^{\text{HS}}$, and the generalized Sait\^{o}-type expression for $\eta_\infty^{\text{HS}}$ \cite{Abade_visco2010},
\begin{equation}\label{eq:etainf_HS}
\frac{\eta_\infty^{\text{HS}}}{\eta_0} = 1 + \frac{5}{2}\phi \frac{1+S}{1 - \phi(1+S)},  
\end{equation}
where $S = 1.001\phi + 0.95\phi^2 - 2.15\phi^3$. Moreover, we use the precise formula \cite{Banchio1999Gsess} for the structure factor peak height
\begin{equation}\label{eq:Smax_HS}
S^{HS}(q_m) \approx 1 + 0.644\phi g^{HS}(x=1^+),
\end{equation}
where $g^{HS}(x=1^+) = (1-0.5\phi)/{(1-\phi)}^3$  is the Carnahan-Starling contact value for the hard-sphere rdf.
For $S^{\text{HS}}(q\to0)$ in \expressionname~\eqref{eq:KD-GSE}, we employ the Carnahan-Starling equation of state \cite{Hansen_McDonald1986}.

In testing the GSE relations in \expressionsname~\eqref{eq:Self_Cage_GSE} and \eqref{eq:KD-GSE} for 
low-salinity systems, for the diffusion properties we use the self-part corrected $\delta\gamma$ scheme, with $d_s$ calculated by the PA scheme.
For $\eta_\infty$ we use the corrected $\delta\gamma$ scheme according to \expressionname~\eqref{eq:etainf_dgCorr}.

\begin{figure}
\begin{center}
\hspace{-1em}
\includegraphics[width=.33\textwidth,angle=-90]{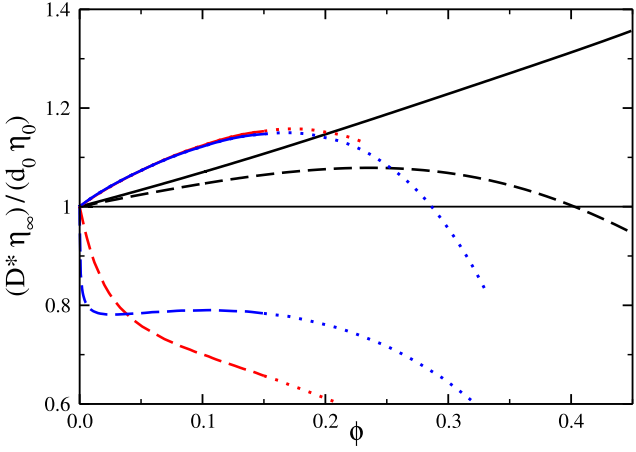}
\end{center}
\vspace{-1em}
\caption{Test of the two GSE relations in \expressionname~\eqref{eq:Self_Cage_GSE}, relating $\eta_\infty$ to $d_s$ and $d_{\text{cge}}$, respectively.
Solid lines: $(d_s/d_0) \times (\eta_\infty/\eta_0)$ vs. $\phi$. 
Dashed lines: $(d_{\text{cge}}/d_0) \times (\eta_\infty/\eta_0)$ vs. $\phi$.
Black curves are hard-sphere results obtained from \expressionsname~\eqref{eq:ds_HS}, \eqref{eq:Hmax_HS}, \eqref{eq:etainf_HS}, and \eqref{eq:Smax_HS}.
Red and blue curves are corrected $\delta\gamma$-scheme results, using as input the MPB-RMSA results for $S(q)$ of
the low-salinity systems CS-1 and CS-2, for $n_s = 10^{-6}$ M.
}
\label{fig:Trans_Cage_GSEs}
\end{figure}

The validity of the two GSE relations in \expressionname~\eqref{eq:Self_Cage_GSE} is examined in \figurename~\ref{fig:Trans_Cage_GSEs}. A valid 
GSE relation is reflected by a horizontal line of unit height. For neutral hard spheres (black lines), the product 
$(d_s^{\text{HS}}/d_0) \times (\eta_\infty^{\text{HS}}/\eta_0)$ is well approximated, for all displayed $\phi$, by its first-order in $\phi$ expansion
given by $1 + 0.67\phi$, showing a more than $20\%$ violation of this GSE relation for $d_s^{\text{HS}}$ when $\phi \gtrsim 0.3$.

Different from $d_s^{\text{HS}}$, the GSE scaling for $d_{\text{cge}}^{\text{HS}}$ is approximately satisfied with a maximal deviation
from one of $8\%$.
Thus, for hard-sphere like colloidal particles available only in amounts too small for a mechanical rheological experiment, one can 
determine $\eta_\infty$ approximately from a dynamic scattering experiment measuring $D(q_m)$.

According to \figurename~\ref{fig:Trans_Cage_GSEs}, in low-salinity systems of charged particles, the GSE-relation for $d_s$ is overall of similar 
accuracy as that for hard spheres, although the deviations from one are larger at smaller $\phi$. The curves for $(d_s/d_0) \times (\eta_\infty/\eta_0)$
obtained for the two low-salinity systems coincide practically for $\phi \lesssim 0.15$. The downturn of these two curves at larger $\phi \gtrsim 0.18$,
indicated by the dotted curve continuations in \figurename~\ref{fig:Trans_Cage_GSEs}, is
not shared by the ASD simulation data (c.f., \figurename~25 in \cite{Banchio2008}). This is an artifact of the PA scheme which,
as discussed already in relation to \figurename~\ref{fig:Hq_varsalt_various_matrix},
tends to underestimate $d_s$ at larger $\phi$. 

Different from neutral hard spheres, in low-salinity systems the GSE relation for $d_{\text{cge}}$ is manifestly violated 
already at very low $\phi$. The strong difference in the $(d_{\text{cge}}/d_0) \times (\eta_\infty/\eta_0)$ curves for the two considered low-salinity
systems is due to the different $\phi$-dependence of their respective $S(q_m)$. The pronounced decline of both curves at low $\phi$ is
mainly triggered by the sharp low-$\phi$ rise of $S(q_m)$ in low-salinity systems.  

\begin{figure}
\begin{center}
\vspace{-1em}
\hspace{-1em}
\includegraphics[width=.37\textwidth,angle=-90]{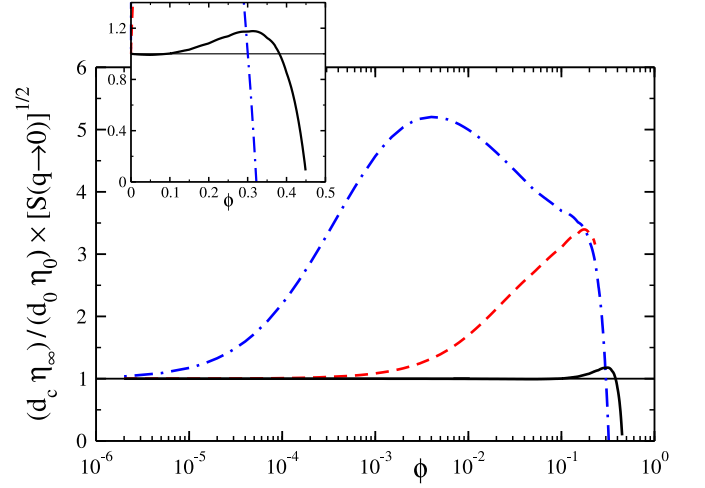}
\end{center}
\vspace{-1em}
\caption{Test of Kholodenko-Douglas GSE relation in \expressionname~\eqref{eq:KD-GSE}.
Black curves: neutral hard-sphere results based on precise analytic expressions for $\eta_\infty^{\text{HS}}$,
$S^{\text{HS}}(q\to0)$, and $K^{\text{HS}}$ calculated in self-part corrected $\delta\gamma$-scheme, with $d_s$ taken from \expressionname~\eqref{eq:ds_HS}.
Red (dashed) and blue (dashed-dotted) curves are self-part corrected $\delta\gamma$-scheme results for the low-salinity systems \mbox{CS-1} and \mbox{CS-2}
(using $n_s = 10^{-6}$ M), respectively, with $d_s$ calculated using the PA scheme, and $S(q)$ using the MPB-RMSA method.
}
\label{fig:KD-GSE}
\end{figure}

Kholodenko and Douglas \cite{Kholodenko1995} have proposed the GSE relation in \expressionname~\eqref{eq:KD-GSE} using mode-coupling theory like
arguments. For neutral spheres at low $\phi$, we can check this relation analytically using the numerically precise second-order
virial expansion results for $K^{\text{HS}}$ in \cite{Cichocki2002}, $\eta_\infty^{\text{HS}}$ in \cite{Cichocki2003},
and $S^{\text{HS}}(q\to0)$ given by the Carnahan-Starling equation of state. This leads to
\begin{equation}\label{eq:KD-GSE_HS_virial}
\frac{d_C^{\text{HS}}}{d_0} \times \frac{\eta_{\infty}^{\text{HS}}}{\eta_0} \times \sqrt{S^{\text{HS}}(q \to 0,\phi)} = 1 - 0.046\phi + 1.3713\phi^2 + \mathcal{O}(\phi^3),
\end{equation}
where the first-order virial coefficient is indeed close to zero. Using the analytic expressions for
$\eta_\infty^{\text{HS}}$ in \expressionname~\eqref{eq:etainf_HS}, 
and $K^{\text{HS}}$ calculated in the self-part corrected $\delta\gamma$-scheme with $d_s$ according to \expressionname~\eqref{eq:ds_HS},
we can test the KD-GSE relation in fact in the full fluid-state regime of neutral spheres. According to the inset in \figurename~\ref{fig:KD-GSE},
the KD-GSE for neutral spheres is valid to decent accuracy up to $\phi \lesssim 0.4$,
with a maximal positive-valued deviation from one at $\phi \approx 0.3$ of less than $18\%$. Strong negative-valued deviations from one
are observed for $\phi \gtrsim 0.4$, where the KD-GSE relation ceases to be applicable to neutral hard spheres.

The corrected $\delta\gamma$-scheme results included in \figurename~\ref{fig:KD-GSE} demonstrate the striking violation of the KD-GSE relation, when low-salinity
systems of charged particles are considered. A clear violation of this relation is observed for all concentrations 
$\phi \gtrsim 10^{-4}$ in the case of the low-salinity system CS-1, and
for $\phi \gtrsim 10^{-2}$ in the case of system CS-2, where, for both systems, a residual salt concentration of $n_s = 10^{-6}$ M has been assumed.
The maximal (positive-valued) violation of the KD-GSE relation occurs roughly at a volume fraction where the $d_c(\phi)$ of charged spheres attains
its maximum. The maximum in $d_c(\phi)$, in turn, is the result of a competition, with increasing $\phi$, between decreasing compressibility and decreasing sedimentation
coefficient. The concentration at the peak of $d_c(\phi)$ is determined roughly from $k^2_c(\phi) = k^2_s$ \cite{HeinenBSA2011}. The downturn of the
$d_c \eta_\infty \sqrt{S(q\to0)} / (d_0 \eta_0)$ curve at large $\phi$, observable in \figurename~\ref{fig:KD-GSE} for neutral and charged spheres alike,
is triggered by the large-$\phi$ decline both of $K$ and $\eta_\infty$.

\section{Conclusions}\label{sec:conclusions}

We have presented a comprehensive theoretical and computer simulation study
of short-time dynamic properties of colloidal spheres, interacting by a HSY pair potential.
In this study, the accuracy of essentially analytic and very fast methods of calculating $H(q)$, $d_s$ and $\eta_\infty$ has been explored
in comparison with numerically expensive ASD simulation results, obtained as functions of volume fraction and added salt content.
We have been particularly concerned with the low screening (low salinity)
and infinite screening (i.e., neutral sphere) limits of the HSY model.

The static pair functions $S(q)$ and $g(r)$, required as the only input to the PA and (self-part corrected) $\delta\gamma$ schemes, have been
determined using our recently developed MPB-RMSA method. The PA scheme, which precisely (and only) accounts for two-body HIs,
is for arbitrary salt concentration in excellent agreement with the simulation data for $H(q)$, provided that $\phi \lesssim 0.1$.
With regard to the high-frequency viscosity, the PA-scheme predictions are in good agreement with the simulation data for volume fractions
up to $\phi \approx 0.1$ for neutral hard spheres, and up to $\phi \approx 0.2$ for strongly charged spheres in the weak screening
(low salinity) regime. At larger $\phi$, three-body and higher-order HIs become influential.

The self-part corrected $\delta\gamma$ scheme for $H(q)$ is
in good agreement with our ASD results for the hydrodynamic function of charged spheres at all considered $\phi$ and $n_s$ values.
When the self-part corrected $\delta\gamma$ scheme is applied to neutral hard spheres, using $d_s^{\text{HS}}$ according to \expressionname~\eqref{eq:ds_HS},
also $H^{\text{HS}}(q)$ is predicted to very good accuracy even up to $\phi \approx 0.4$, including the small-$q$ region.

We have shown that the (second-order) $\delta\gamma$-scheme expression for $\eta_\infty$, given by \expressionsname~\eqref{eq:etainf_deltagamma}-\eqref{eq:lambda2},
is only applicable to hard-sphere like systems (i.e., at large screening). Its predictions for $\eta_\infty^{\text{HS}}$, however, agree 
well with the ASD data up to $\phi \lesssim 0.4$. At larger $\phi$, it underestimates the high-frequency viscosity. Based on
arguments applying to low-salinity charge-stabilized systems only, we have introduced in \expressionname~\eqref{eq:etainf_dgCorr} a simple correction to
the $\delta\gamma$-scheme result for $\eta_\infty$. This corrected $\delta\gamma$ scheme predicts to good accuracy the high-frequency 
viscosity of low-salinity systems, even up to the freezing concentration.
Different from the self-part corrected $\delta\gamma$ scheme for $H(q)$ introduced in \expressionname~\eqref{eq:Hq_dgCorr}, which applies at any salt
concentration under the condition that $\phi \lesssim 0.15$, the corrected $\delta\gamma$ scheme for $\eta_\infty$ in \expressionname~\eqref{eq:etainf_dgCorr}
is valid at low salinities only. An appropriately corrected
$\delta\gamma$ scheme for $\eta_\infty$, applicable for arbitrary salinities, is still missing. Its development could be the topic of a future study.      

An interesting application of the hybrid $\delta\gamma$ schemes for $H(q)$ and $\eta_\infty$ has been our validity tests of three approximate
generalized Stokes-Einstein relations linking $\eta_\infty$ to $d_s$, $d_{\text{cge}}$, and $d_c$, respectively. For an optimal test of these
relations in the special case of neutral hard spheres, precise analytic expressions for $d_s^{\text{HS}}$, $H^{\text{HS}}(q_m)$, $S^{\text{HS}}(q_m)$, and
$S^{\text{HS}}(q\to0)$ have been used, valid in the whole fluid-state concentration range. The key finding from our validity tests of GSE
relations is the strong dependence of their accuracies on the range and character of the particle interactions. The most striking example in case
is the Kholodenko-Douglas GSE relation, which applies decently well to neutral spheres up to $\phi \approx 0.4$. The very same relation, however,
is strongly violated in low-salinity suspensions already at very low volume fractions.

\begin{acknowledgments}
M.H. acknowledges support by the International Helmholtz Research School of
Biophysics and Soft Matter (IHRS BioSoft).
A.J.B. acknowledges financial support from SeCyT-UNC and CONICET.
This work was under appropriation of funds from
the Deutsche Forschungsgemeinschaft (SFB-TR6, project B2). 
\end{acknowledgments}


\begin{thebibliography}{10}

\bibitem{Russel1989}
W.~B. Russel, D.~A. Saville, and W.~R. Schowalter.
\newblock {\em Colloidal Dispersions}.
\newblock Cambridge University Press, Cambridge, 1989.

\bibitem{RiesePRL2000}
D.~O. Riese, G.~H. Wegdam, W.~L. Vos, R.~Sprik, D.~Fenistein, J.~H.~H.
  Bongaerts, and G.~Gr\"{u}bel.
\newblock {\em {Phys. Rev. Lett.}}, {85}:{5460--5463}, {2000}.

\bibitem{HolmqvistPRL2010}
P.~Holmqvist and G.~N\"{a}gele.
\newblock {\em {Phys. Rev. Lett.}}, {104}:{058301}, {2010}.

\bibitem{Patkowski2005}
J.~Gapinski, A.~Wilk, A.~Patkowski, W.~H\"{a}ussler, A.~J. Banchio, R.~Pecora,
  and G.~N\"{a}gele.
\newblock {\em {J. Chem. Phys.}}, {123}:{054708}, {2005}.

\bibitem{HeinenBSA2011}
M.~Heinen, F.~Zanini, F.~Roosen-Runge, D.~Fedunov\'a, F.~Zhang, M.~Hennig,
  Seydel T., R.~Schweins, M.~Sztucki, M.~Antal\'ik, F.~Schreiber, and
  G.~N\"agele.
\newblock submitted.

\bibitem{Roosen-Runge2011}
F.~Roosen-Runge, M.~Hennig, F.~Zhang, R.~M.~J. Jacobs, M.~Sztucki, H.~Schober,
  T.~Seydel, and F.~Schreiber.
\newblock {\em {PNAS}}, {108}:{11815--11820}, {2011}.

\bibitem{Verwey_Overbeek1948}
E.~J.~W. Verwey and J.~T.~G. Overbeek.
\newblock {\em Theory of the Stability of Lyophobic Colloids}.
\newblock Elsevier, New York, 1948.

\bibitem{Pusey1991}
P.~N. Pusey.
\newblock {\em Liquids, Freezing and the Glass Transition}.
\newblock Elsevier, Amsterdam, 1991.

\bibitem{heinen2010short}
M.~Heinen, P.~Holmqvist, A.~J. Banchio, and G.~N\"{a}gele.
\newblock {\em {J. Appl. Crystallogr.}}, {43}:{970--980}, {2010}.

\bibitem{BanchioPRL2006}
A.~J. Banchio, J.~Gapinski, A.~Patkowski, W.~H\"{a}ussler, A.~Fluerasu,
  S.~Sacanna, P.~Holmqvist, G.~Meier, M.~P. Lettinga, and G.~N\"{a}gele.
\newblock {\em {Phys. Rev. Lett.}}, {96}:{138303}, {2006}.

\bibitem{Farago2003}
W.~H\"{a}ussler and B.~Farago.
\newblock {\em {J. Phys.-Cond. Matt.}}, {15}:{S197--S204}, {2003}.

\bibitem{Longeville2008}
C.~Le~Coeur and S.~Longeville.
\newblock {\em {Chem. Phys.}}, {345}:{298--304}, {2008}.

\bibitem{Mason2010}
T.~M. Squires and T.~G. Mason.
\newblock {\em {Annu. Rev. Fluid Mech.}}, {42}:{413--438}, {2010}.

\bibitem{Banchio2008}
A.~J. Banchio and G.~N\"{a}gele.
\newblock {\em {J. Chem. Phys.}}, {128}:{104903}, {2008}.

\bibitem{Mazur1984}
C.~W.~J. Beenakker and P.~Mazur.
\newblock {\em {Physica A}}, {126}:{349--370}, {1984}.

\bibitem{Beenakker1983}
C.~W.~J. Beenakker and P.~Mazur.
\newblock {\em {Physica A}}, {120}:{388--410}, {1983}.

\bibitem{Beenakker1984}
C.~W.~J. Beenakker.
\newblock {\em {Physica A}}, {128}:{48--81}, {1984}.

\bibitem{Heinen2011}
M.~Heinen, P.~Holmqvist, A.~J. Banchio, and G.~N\"{a}gele.
\newblock {\em {J. Chem. Phys.}}, {134}:{044532}, {2011}.

\bibitem{Heinen_Erratum_2011}
M.~Heinen, P.~Holmqvist, A.~J. Banchio, and G.~N\"{a}gele.
\newblock {\em {J. Chem. Phys.}}, {134}:{129901}, {2011}.

\bibitem{Russel1981}
W.~B. Russel and D.~W. Benzing.
\newblock {\em {J. Colloid Interface Sci.}}, {83}:{163--177}, {1981}.

\bibitem{Denton2000}
A.~R. Denton.
\newblock {\em {Phys. Rev. E}}, {62}:{3855--3864}, {2000}.

\bibitem{Hansen_McDonald1986}
J.-P. Hansen and I.~R. McDonald.
\newblock {\em Theory of Simple Liquids}.
\newblock Academic Press, London, 2 edition, 1986.

\bibitem{Snook1992}
I.~K. Snook and J.~B. Hayter.
\newblock {\em {Langmuir}}, {8}:{2880--2884}, {1992}.

\bibitem{Hansen1982}
J.-P. Hansen and J.~B. Hayter.
\newblock {\em {Mol. Phys.}}, {46}:{651--656}, {1982}.

\bibitem{Rogers1984}
F.~J. Rogers and D.~A. Young.
\newblock {\em {Phys. Rev. A}}, {30}:{999--1007}, {1984}.

\bibitem{Percus1958}
J.~K. Percus and G.~J. Yevick.
\newblock {\em {Phys. Rev.}}, {110}:{1--13}, {1958}.

\bibitem{Wertheim1963}
M.~S. Wertheim.
\newblock {\em {Phys. Rev. Lett.}}, {10}:{321--323}, {1963}.

\bibitem{Nagele1996}
G.~N\"{a}gele.
\newblock {\em {Phys. Rep.}}, {272}:{216--372}, {1996}.

\bibitem{Dhont1996}
J.~K.~G. Dhont.
\newblock {\em An Introduction to Dynamics of Colloids}.
\newblock Elsevier, Amsterdam, 1996.

\bibitem{Jones1991}
R.~B. Jones and P.~N. Pusey.
\newblock {\em {Annu. Rev. Phys. Chem.}}, {42}:{137--169}, {1991}.

\bibitem{Cichocki2008}
P.~Szymczak and B.~Cichocki.
\newblock {\em {J. Stat. Mech.-Theory Exp.}}, page {P01025}, {2008}.

\bibitem{BanchioBrady:03}
A.~J. Banchio and J.~F. Brady.
\newblock {\em {J. Chem. Phys.}}, {118}:{10323}, {2003}.

\bibitem{AbadeJCP2010}
G.~C. Abade, B.~Cichocki, M.~L. Ekiel-Je\.{z}ewska, G.~N\"{a}gele, and
  E.~Wajnryb.
\newblock {\em {J. Chem. Phys.}}, {132}:{014503}, {2010}.

\bibitem{Abade_visco2010}
G.~C. Abade, B.~Cichocki, M.~L. Ekiel-Je\.{z}ewska, G.~N\"{a}gele, and
  E.~Wajnryb.
\newblock {\em {J. Chem. Phys.}}, {133}:{084906}, {2010}.

\bibitem{jeff_oni:84}
D.~J. Jeffrey and Y.~Onishi.
\newblock {\em {J. Fluid Mech.}}, {139}:{261--290}, {1984}.

\bibitem{Schmitz1988}
R.~B. Jones and R.~Schmitz.
\newblock {\em {Physica A}}, {149}:{373--394}, {1988}.

\bibitem{Batchelor1972}
G.~K. Batchelor and J.~T. Green.
\newblock {\em J. Fluid Mech.}, 56:401--427, 1972.

\bibitem{Russel1984}
W.~B. Russel.
\newblock {\em {J. Chem. Soc., Faraday Trans.}}, {80}:{31--41}, {1984}.

\bibitem{Cichocki1999}
B~Cichocki, ML~Ekiel-Jezewska, and E~Wajnryb.
\newblock {\em {J. Chem. Phys.}}, {111}:{3265--3273}, {1999}.

\bibitem{Cichocki2002}
B.~Cichocki, M.~L. Ekiel-Jezewska, P.~Szymczak, and E.~Wajnryb.
\newblock {\em {J. Chem. Phys.}}, {117}:{1231--1241}, {2002}.

\bibitem{Cichocki2003}
B.~Cichocki, M.~L. Ekiel-Jezewska, and E.~Wajnryb.
\newblock {\em {J. Chem. Phys.}}, {119}:{606--619}, {2003}.

\bibitem{Genz1991}
U.~Genz and R.~Klein.
\newblock {\em {Physica A}}, {171}:{26--42}, {1991}.

\bibitem{Gapinski2006}
A.~J. Banchio, J.~Gapinski, A.~Patkowski, W.~H\"{a}ussler, A.~Fluerasu,
  S.~Sacanna, P.~Holmqvist, G.~Meier, M.~P. Lettinga, and G.~N\"{a}gele.
\newblock {\em {Phys. Rev. Lett.}}, {96}:{138303}, {2006}.

\bibitem{Sinn1999}
E.~Overbeck, C.~Sinn, and M.~Watzlawek.
\newblock {\em {Phys. Rev. E}}, {60}:{1936--1939}, {1999}.

\bibitem{Nagele1994}
G.~N\"{a}gele, B.~Steininger, U.~Genz, and R.~Klein.
\newblock {\em {Phys. Scr.}}, {55}:{119--126}, {1994}.

\bibitem{Nagele1995}
G.~N\"{a}gele, B.~Mandl, and R.~Klein, {1995}.

\bibitem{Watzlawek1999}
M.~Watzlawek and G.~N\"{a}gele.
\newblock {\em {J. Colloid Interface Sci.}}, {214}:{170--179}, {1999}.

\bibitem{Abade2011}
G.~C. Abade, B.~Cichocki, M.~L. Ekiel-Je\.{z}ewska, G.~N\"{a}gele, and
  E.~Wajnryb.
\newblock {\em {J. Chem. Phys.}}, {134}:{244903}, {2011}.

\bibitem{Bergenholtz2000}
F.~M. Horn, W.~Richtering, J.~Bergenholtz, N.~Willenbacher, and N.~J. Wagner.
\newblock {\em {J. Coll. Interf. Sci.}}, {225}:{166--178}, {2000}.

\bibitem{Ladd:90}
A.~J.~C. Ladd.
\newblock {\em {J. Chem. Phys.}}, {93}:{3484}, {1990}.

\bibitem{MoSangani1994}
G.~Mo and A.~S. Sangani.
\newblock {\em {Phys. Fluids}}, {6}:{1637--1652}, {1994}.

\bibitem{SegreBehrend:95}
P.~N. Segr\`e, O.~P. Behrend, and P.~N. Pusey.
\newblock {\em {Phys. Rev. E}}, {52}:{5070}, {1995}.

\bibitem{Ladd:95}
A.J.C Ladd, H.~Gang, Zhu~J. X., and D.~A. Weitz.
\newblock {\em {Phys. Rev. E}}, {52}:{6550}, {1995}.

\bibitem{StevensRobbins1993}
M.~J. Stevens and M.~O. Robbins.
\newblock {\em {J. Chem. Phys.}}, {98}:{2319--2324}, {1993}.

\bibitem{HansenVerlet1969}
J.-P. Hansen and L.~Verlet.
\newblock {\em {Phys. Rev.}}, {184}:{151}, {1969}.

\bibitem{KremerGrest1986}
K.~Kremer, M.~O. Robbins, and G.~S. Grest.
\newblock {\em {Phys. Rev. Lett.}}, {57}:{2694--2697}, {1986}.

\bibitem{Gapinski2010}
J.~Gapinski, A.~Patkowski, and G.~N\"{a}gele.
\newblock {\em {J. Chem. Phys.}}, {132}:{054510}, {2010}.

\bibitem{Holmqvist2011}
M.~Heinen, P.~Holmqvist, A.~J. Banchio, and G.~N\"{a}gele.
\newblock Article in preparation.

\bibitem{Banchio1999Gsess}
A.~J. Banchio, G.~N\"{a}gele, and J.~Bergenholtz.
\newblock {\em {J. Chem. Phys.}}, {111}:{8721--8740}, {1999}.

\bibitem{Kholodenko1995}
A.~L. Kholodenko and J.~F. Douglas.
\newblock {\em {Phys. Rev. E}}, {51}:{1081--1090}, {1995}.

\bibitem{Gaigalas1995}
A.~K. Gaigalas, V.~Reipa, J.~B. Hubbard, J.~Edwards, and J.~Douglas.
\newblock {\em {Chem. Eng. Sci.}}, {50}:{1107--1114}, {1995}.

\bibitem{Cohen1998}
D.~E. Cohen, G.~M. Thurston, R.~A. Chamberlin, G.~B. Benedek, and M.~C. Carey.
\newblock {\em {Biochemistry}}, {37}:{14798--14814}, {1998}.

\bibitem{Boogerd2001}
P.~Boogerd, B.~Scarlett, and R.~Brouwer.
\newblock {\em {Irrig. Drain.}}, {50}:{109--128}, {2001}.

\bibitem{Nettesheim2008}
F.~Nettesheim, M.~W. Liberatore, T.~K. Hodgdon, N.~J. Wagner, E.~W. Kaler, and
  M.~Vethamuthu.
\newblock {\em {Langmuir}}, {24}:{7718--7726}, {2008}.

\end{thebibliography}
\end{document}